\documentclass[twocolumn,dvipsnames,nofootinbib,floatfix,groupedaddress,superscriptaddress]{revtex4-1}
\PassOptionsToPackage{unicode}{hyperref}
\PassOptionsToPackage{bookmarks=false}{hyperref}
\newcommand{\coll}{{\rm coll}}
\newcommand{\width}{{\rm fwhm}}
\newcommand{\nux}{\ensuremath{{\bar\nu_x}}}

\newcommand{\sph}{{\rm sph}}
\newcommand{\MeV}{\,\unit{MeV}}
\newcommand{\km}{\,\unit{km}}

\newcommand{\mfp}{\mathrm{mfp}}
\newcommand{\eff}{\mathrm{\rm eff}}
\newcommand{\res}{\mathrm{res}}
\newcommand{\osc}{\mathrm{osc}}
\newcommand{\tpb}{t_{\rm pb}}
\newcommand{\tot}{\text{tot}}

\newcommand{\parfrac}[2]{\left(\frac{#1}{#2}\right)}
\usepackage[utf8]{inputenc}
\usepackage{xcolor,soul}
\usepackage{units}
\usepackage{siunitx}
\usepackage{amsmath}
\usepackage{amssymb,amsthm,amsfonts} 
\usepackage{bbold}
\usepackage{paralist}
\usepackage{xcolor}
\usepackage{graphicx}
\usepackage[T1]{fontenc}
\usepackage{braket} 
\usepackage[normalem]{ulem}

\usepackage{slashed} \hfuzz=2pt \usepackage{lmodern}
\usepackage{hyperref}
\begin{document}

\title{Resonance production of keV sterile neutrinos in core-collapse supernovae\newline and lepton number diffusion}
  
\author{Vsevolod Syvolap}
\affiliation{Discovery Center, Niels Bohr Institute, Copenhagen
      University, Blegdamsvej 17, Copenhagen, DK-2100, Denmark}
\affiliation{Taras Shevchenko National Technical University, Kiev,
      Ukraine }

\author{Oleg Ruchayskiy}
\affiliation{Discovery Center, Niels Bohr Institute, Copenhagen
      University, Blegdamsvej 17, Copenhagen, DK-2100, Denmark}
      
\author{Alexey Boyarsky}
\affiliation{Lorentz Institute, Leiden University, Niels Bohrweg 2, Leiden,
  NL-2333 CA, The Netherlands}

\begin{abstract}
   We investigate how hypothetical particles -- sterile neutrinos -- can be produced in the interior of exploding supernovae via the resonant conversion of $\bar\nu_\mu$ and $\bar \nu_\tau$.
  The novelty of our treatment lies in the proper account of the resulting lepton number diffusion.
  We compute the yield of sterile neutrinos and find that even after taking into account back reaction, sterile neutrinos can carry out a sizeable fraction of the total energy of the explosion comparable to that of active neutrinos.
  \textcolor{black}{The production is, however, sensitive to the temperature in the inner supernovae regions, making robust predictions of challenging.}
  In order to understand whether this production affects supernova evolution and can therefore be constrained, detailed simulations including the effects of sterile neutrinos are needed.

\end{abstract}
\maketitle

\section{Introduction and outlook}

Exploding supernovae (SNe) are characterized by high temperatures
$T\sim \mathcal{O}(10)$~MeV and high densities of baryons.  
This makes them unique laboratories that can copiously produce hypothetical feebly interacting particles~\cite{Raffelt:1990yz,Raffelt:1996wa,Raffelt:1999tx}, including axions, dark photons, millicharged particles, sterile neutrinos (see \textit{e.g.}~\cite{Essig:2013lka,Alekhin:2015byh}).

SN medium is
not transparent for neutrinos of all flavors, and their dispersion relations
change, as compared to the vacuum case $\omega = |k|$~\cite{Notzold:1987ik}.
In the models with \textit{sterile neutrinos} ($\nu_s$) -- massive neutral particles,
that mix with active neutrinos -- this may lead to the enhancement of
active-sterile mixing, similarly to the solar MSW
effect~\cite{Wolfenstein:1977ue,Mikheev:1986gs}.  
Feeble interaction of the
resulting particles allow them to escape from the interiors of SNe.

The question of sterile neutrino production during supernovae explosion, their
effects on explosion, and the stellar nucleosynthesis has been studied in
the past~\cite{Kainulainen:1990bn,Raffelt:1992bs,Peltoniemi1992,Shi:1993ee,Kusenko:1997sp,Dolgov:2000pj,Dolgov:2000ew,Dolgov:2000jw,Abazajian:2001nj,Fuller:2003gy,Barkovich:2004jp,Hidaka:2006sg,Hidaka:2007se,Kusenko:2008gh,Fuller:2009zz,Raffelt:2011nc,Wu:2013gxa,Warren:2014qza,Warren:2016slz,Rembiasz:2018lok}.
These studies mostly concentrated on the mixing of the sterile neutrino with
electron flavor, owing to the presence of the significant electron lepton
number $L_e$ in the supernova.  The production of $\nu_s$ from $\mu$ and
$\tau$ flavors has been considered in \cite{Abazajian:2001nj,Raffelt:2011nc,Zhou:2015jha}.\footnote{In what
  follows we will use the notation $\nu_x$ to denote collectively
  $(\nu_{\mu},\nu_{\tau})$ and $\bar{\nu}_x$ for
  $(\bar{\nu}_{\mu}, \bar{\nu}_{\tau})$ respectively.}  
These works took into account production via scattering in the constant-density core of the supernova,
expecting that the effect should be the strongest there due to the high density of matter and temperature. 

The question of production of $\nu_s$, mixed with $\nu_x$ has been re-analysed recently in \cite{Arguelles:2016uwb} where it had been noticed that outside the core the resonant MSW-like conversion of $\bar\nu_x$ into sterile neutrino $\nu_s$ was possible (see also~\cite{Suliga:2019bsq}).
It was argued in \cite{Arguelles:2016uwb} that such a conversion can be quite efficient and can lead to a significant flux of $\nu_s$ for mixing angles as small as $\sin^2\bigl(2\theta_{\mu,\tau}\bigr) \sim 10^{-12}$.

In this work we re-analyze sterile neutrino production in the course of
supernovae explosion, taking into account the back-reaction of sterile neutrino
emission on the local density of anti-neutrinos. We demonstrate that
\begin{compactitem}[--]
\item the local density of anti-neutrinos $\bar\nu_x$ in the resonance zone is quickly reduced (the chemical potential $\mu_x \gtrsim T$ is generated), thus slowing the sterile neutrino production.
\item The diffusion processes are not efficient enough to restore the population of $\bar\nu_x$ in the resonance zone.
\item The exact amount of energy carried by sterile neutrinos is sensitive to the temperature in the inner SN regions.  This makes robust predictions of sterile neutrino flux challenging, as these temperatures are not sufficiently constrained.
\end{compactitem}
As a result the process of sterile neutrino production eventually switches off.  
Nevertheless, we find that sterile neutrinos can carry out a significant fraction of the total
energy of the explosion, comparable with the energy flux of a flavor of active neutrinos.
\emph{This constitutes the main result of our paper.}

\bigskip

\noindent
The structure of the paper and the main points of each Section are as follows:
\begin{compactitem}[$\bullet$]
\item Section~\ref{sec:sketch} lists the formulas that are sufficient to reproduce our results and explains basic ingredients that enter the computations. Details and comments, accompanying these formulas are provided in Appendices
\item Section~\ref{sec:results} presents our results: we estimate the amount of energy carried away by $\nu_s$, calculate their spectra and evolution of the chemical potential of $\mu$ and $\tau$ flavors in space and time. \textbf{Our main results are summarised in Figs.~\ref{fig:EmittedEnergiesFullDiffusion}.}

\item In Section~\ref{sec:conclusion} we conclude that although sterile neutrino production can be quite efficient, it is difficult to obtain robust constraints on sterile neutrino parameters based on the scarce data we have and that one needs holistic simulations of SN explosions, including sterile neutrinos to see whether too much energy gets carried away through this channel.
\item Appendices~\ref{app:SNmodel}--\ref{asymmetryappendix} provide background information and additional cross-checks; details of the derivation of the kinetic equation; treatment of the diffusion, etc. 
\end{compactitem}

\bigskip
 \textbf{Note added.}

 When this manuscript was finished, the paper~\cite{Suliga:2019bsq} appeared that also investigates the production of $\nu_s$ mixed with $\nu_\tau$ in the SN interior. 
Ref.~\cite{Suliga:2019bsq} analyses the evolution of the lepton asymmetry $Y_\tau$ due to the resonance conversion and the collisional production as well as the feedback on the effective potential. 
The main difference for the resonance conversion study is that we account for the neutrino's lepton number diffusion which eases the back-reaction. Therefore our results are qualitatively similar, the difference can be attributed to different SN explosion models. 
\section{Sketch of the computations}
\label{sec:sketch}
In order to keep the presentation simple and spare readers from technical details, we start by summarising the main steps of our calculations and basic formulas that would allow one to reproduce our results.
Details of the derivation and calculation are provided in Appendix~\ref{ResConversion} below.

In order to compute the production of sterile neutrinos we need to solve a \emph{system of coupled equations}
\begin{enumerate}
\item First equation (Eq.~\eqref{kinetic_res} below) describes the temporal evolution of the distribution function of sterile neutrinos, based on which one can compute, \textit{e.g.}, sterile neutrino energy flux.
\item Second equation (Eq.~\eqref{eq:Ytau} below) governs the evolution of the chemical potential $\mu_x(r,t)$, that describes the back-reaction of the sterile neutrino production on the population of active anti-neutrinos.
\end{enumerate}

The number of $\nu_s$ with energy $E$, resonantly produced by the time $t$ and travelling into the solid angle  $d\Omega$ is
given by (we assume that $E \approx |\vec p|$, \textit{i.e.}\  sterile neutrinos are  ultra-relativistic):
\begin{widetext}
  \begin{equation}
    \boxed{\frac{d^2N_s(t,E)}{dE\,d\Omega} = \int_{0}^t 4\pi R_{\res}^2(E)E^2 \bar
      f^{\rm out}_x\bigl(t',R_{\res}(E),E\bigr)P_{x\to s}(E)e^{-R_{\width}/\lambda_{\mfp}}dt'\,.}
    \label{kinetic_res}
  \end{equation}
\end{widetext}
Expression~\eqref{kinetic_res} requires several comments.
$R_{\res}(E)$ is the radius, at which resonance condition is satisfied for anti-neutrinos with the energy $E$.
Relation $r=R_{\res}(E)$ can be inverted to form $E=E_{\res}(r)$ and determines the value of the energy of $\nu_s$ produced at radius~$r$:
\begin{equation}
  E_{\res}(r)=\frac{m_s^2}{V_\eff(r)}.
  \label{ResonanceEnergy_res}
\end{equation}
$V_\eff(r)$ is the \emph{effective potential} of  anti-neutrinos~\cite{Notzold:1987ik}.
For the $\bar\nu_\mu$:
\begin{equation}
  V_{\eff}(r) = -\frac{G_F}{\sqrt{2}}N_{\rm b}\Bigl(Y_n - 2Y_{\nu_e} - 2Y_{\nu_\tau} - 4Y_{\nu_\mu}-2Y_{\mu}\Bigr)
  \label{Veff}
\end{equation}
Here $Y_{i} \equiv \frac{N_{i} - N_{\bar\imath}}{N_{\rm b}}$ is the asymmetry in $i^{th}$ particle ($i = \{n, p, e,\mu,\tau,\nu_{e},\nu_{\mu},\nu_{\tau}\}$) , $N_{\rm b}$ is the baryons number density.
All these quantities are functions of position, see App.~\ref{app:SNmodel}.
The effective potential for $\bar\nu_\tau$ is obtained by the replacement $\mu \leftrightarrow \tau$ and $ \nu_{\mu} \leftrightarrow \nu_{\tau}$ in~\eqref{Veff}.
The baryon density $N_{\rm b}$ and asymmetries reach their maximal values in the SN core.
Therefore, the energy, entering~\eqref{kinetic_res} has a minimal value and the spectrum of emitted sterile neutrinos is cut at low energies.

Initial values of asymmetries $Y_i$ we use in the SN model (see Appendix~\ref{app:SNmodel}) are such, that the potential~\eqref{Veff} is \emph{negative}, meaning that the resonance occurs for anti-neutrinos.
Moreover, we found that the potential will not change its sign during the production phase and hence we do not consider any equation for neutrinos conversion.

Numerically, the resonance energy~\eqref{ResonanceEnergy_res} is given by
\begin{equation}
\label{eq:Eres2}
 E_{\text{res}} \sim \unit[9]{MeV} \cdot \left(\frac{m_s}{10 \text{ keV}}\right)^2\cdot\frac{\rho_B}{\unit[3\cdot 10^{14}]{g/cm^3}}.
\end{equation}
\textcolor{black}{where we used for estimate $Y_e = 0.3,  Y_{\nu_e} = 0.1$ and $Y_{\mu} = Y_{\nu_\mu} = Y_{\nu_\mu} = 0$.}

The transition probability $P_{x\rightarrow s}$ is defined as:
\begin{equation}
\label{eq:Px_s}
    P_{x\rightarrow s} = 1-\exp\left[- \frac{\pi^2}{2}\frac{R_{\width}}{L_{\osc}}\right]
\end{equation}
where $R_{\width}$ is the width of the resonance region,
\begin{equation}
  \label{eq:Rwidth}
  R_{\width} = \frac{2 \sin2\theta_0}{\left|\frac{\partial \log V_{\eff}^\res}{\partial r}\right|}\;,
\end{equation}
(derivative of $V_\eff$ is evaluated at $r=R_\res$) and $L_{\osc}$ is the oscillation length at the resonance
\begin{equation}
  \label{eq:Losc}
  L_{\osc} = \frac{2\pi} { |V_{\eff}^{\res}| \sin 2\theta_0}.
\end{equation}
The angle $\theta_0$ is the vacuum active-sterile neutrino mixing and all equations are derived for $\theta_0 \ll 1$.
The resonance is effective when $R_{\width}\gtrsim L_{\osc}$, this ratio is $\propto \sin^2(2\theta_0)$.

The distribution function $\bar f^{\rm out}_x$ describes \emph{outgoing} anti-neutrinos at the radius $r = R_{\res}(E)$.
This function has the \textit{equilibrium} form
\begin{equation}
 \label{eq:FD_res}
  \bar f_{x}(t,r,E) = \frac{1}{(2\pi)^3}\frac{1}{\exp\left[\frac{E + \mu_x(r,t)}{T(r)}\right]+1}
\end{equation}

The evolution of the anti-neutrino population is fully encoded in the chemical potential $\mu_x(r,t)$, we do not take into account temperature evolution during the first second of explosion. 

Factor $e^{-R_{\width}/\lambda_{\mfp}}$  where $\lambda_\mfp$ is the mean free path of $\nux$, streaming radially outwards in the resonance region, accounts for the \textit{neutrino damping} \cite{Stodolskiy1987}, see Section~\ref{sec:damping} below.

For the distribution~\eqref{eq:FD_res} the relation between the chemical potential and  the asymmetry $Y_x$ is
defined as:
\begin{equation}
    \label{eq:Ytau}
    Y_{x} = \frac{1}{N_{\rm b}}\left(\frac{\mu_{x} T^2}{6} + \frac{\mu_{x}^3}{6\pi^2}\right)
  \end{equation}
and the evolution of $Y_x$ is given by the equation

\begin{widetext}
\textcolor{black}{
  \begin{equation}
    \boxed{\frac{\partial Y_{x}(r,t)}{\partial t}= \frac{\pi}{6} \frac{N_{\rm b}(r)}{ G_F^2 r^2}\frac{\partial}{\partial r}\left(\frac{r^2}{N_{\rm b}(r)}\frac{\partial \mu_x(r,t)}{\partial r} \right) + \frac{\pi}{N_{\rm b}(r)} E_{\res}^2(r,t) \bar{f_x}(E_{\res}(r),r,t)P_{x\to s}(E_{\res}(r),r,t) \frac{dE_{\rm res}}{dr}(r,t)\;,}
    \label{diffsimplified_res}
  \end{equation}}
  \end{widetext}
where the first term describes the diffusion of the lepton number and the second term -- the change of lepton asymmetry due to the conversion of anti-neutrinos into~$\nu_s$. \footnote{Notice, that this expression was obtained without any assumption about the value of the chemical potential and is valid for the case of $\mu_x \gg T$ as well.}

Taking into account an implicit dependence of $E_\res$ on $\mu_x$, we can solve~\eqref{diffsimplified_res} for $\mu_x(r,t)$, plug it into Eq.~\eqref{kinetic_res}, and find the distribution function of sterile neutrinos $N_s(E,t)$.

\section{Results}
\label{sec:results}

\subsubsection{Energy output in sterile neutrinos}
\label{sec:Es-output}

The approach sketched in Section~\ref{sec:sketch} allows us to calculate the energy spectra and the total energy emitted in the form of sterile neutrinos $\nu_s$ during the \textit{first second} after the core bounce.\footnote{After $\sim 1$~sec post-bounce the temperature of the area of intense resonance conversion ($r \eqsim 10-20$ km) decreases significantly and the production of sterile neutrinos is essentially switched off. Note, that the temperature inside the core can still be high so this switch off may be less relevant for the collisional production }
Our results are summarised in Fig.~\ref{fig:EmittedEnergiesFullDiffusion} (energy carried out as a function of sterile neutrino parameters).
Fig.~\ref{fig:EmittedEnergiesFullDiffusion} both summarises the production within our fiducial model and demonstrates the level of uncertainties that we associate with such production (see explanation below).  
Section~\ref{sec:conclusion} further discusses the uncertainties and how they influence our ability to constrain particle physics models.

We stress that Fig.~\ref{fig:EmittedEnergiesFullDiffusion} \emph{does not correspond to any constraints on sterile neutrino parameters}.
Given our current knowledge about SN explosions in general and about SN1987A in particular, it is impossible to determine what energy loss would be incompatible with existing scarce observations (see Section~\protect\ref{sec:conclusion} for discussion).

\subsubsection{Qualitative explanation of the results}
\label{sec:qualitative}

\begin{figure}[!t]
  \centering
  \includegraphics[width=\linewidth]{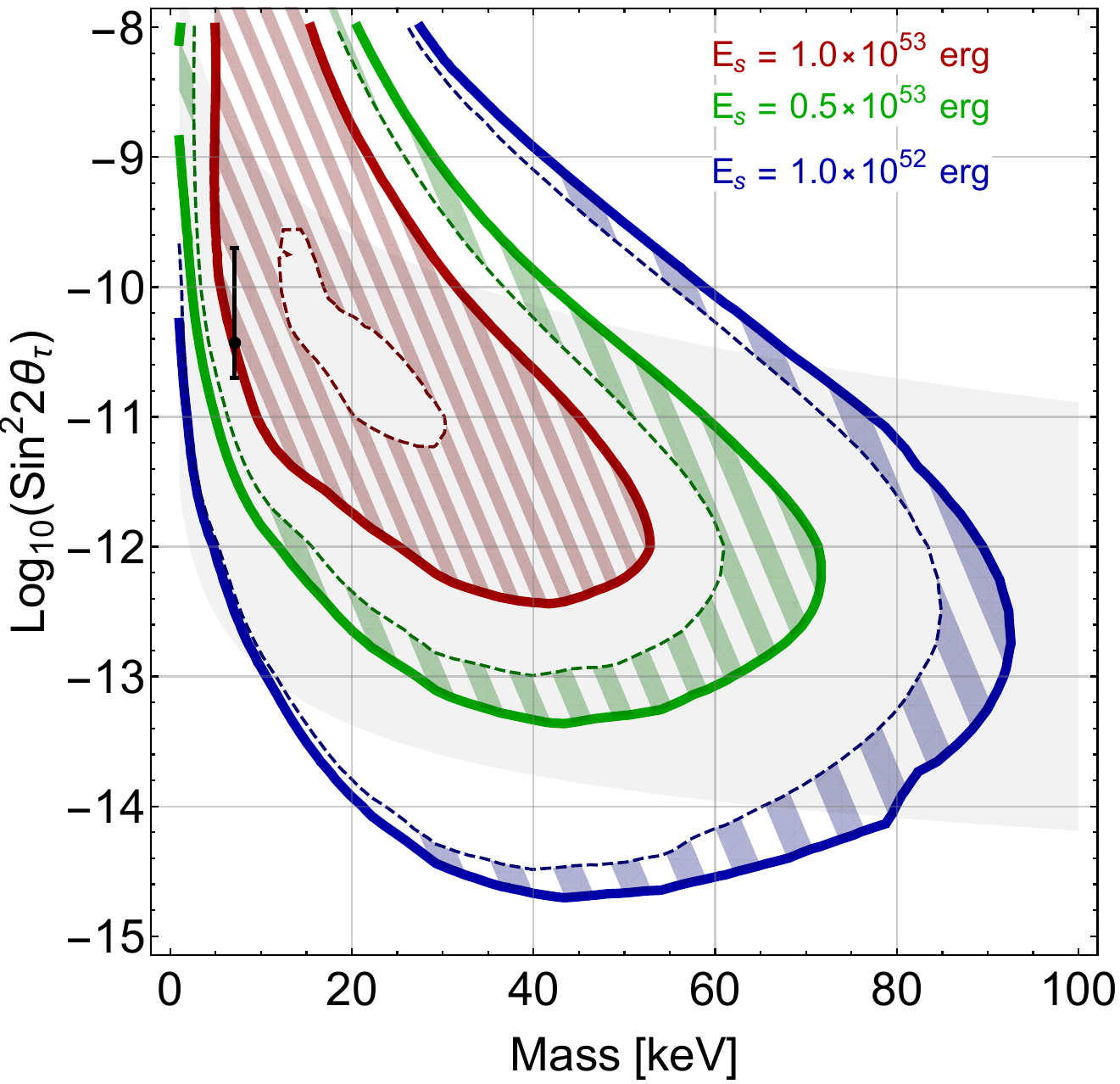}
  \caption[Energy emitted in sterile neutrinos]{\textbf{Main result:}   Energy emitted by sterile neutrinos mixed solely with $\nu_\tau$, produced via resonant conversion during the first second after the SN core bounce (thick solid lines). Thinner dashed lines correspond to same emitted energy in the modified model where the temperature is 10 \% lower.  
  Note, that  the contours with energy output $E_{s} \sim 10^{53}$~erg are only indicative, as we did not include in our treatment the energy loss and cooling due to sterile neutrinos.  Although our analysis did not assume that sterile neutrinos are dark matter particles,  we over impose light grey region to indicate where the correct dark matter abundance of sterile neutrinos can be generated in the \emph{Neutrino Minimal Standard Model} ($\nu$MSM see Section~\protect\ref{sec:sterile-neutrino-DM}).  Black dot with error bars corresponds to the $3.5$~keV signal of~\cite{Boyarsky:2014jta,Bulbul:2014sua} interpreted in terms of decays of sterile neutrino dark matter.}
  \label{fig:EmittedEnergiesFullDiffusion}
\end{figure}

We start with outlining the results and explaining qualitatively the features of the contours in Fig.~\ref{fig:EmittedEnergiesFullDiffusion}.
The parameters of sterile neutrinos are constrained by estimating the amount of energy they may carry away (see Section~\ref{sec:conclusion} for further details).
This energy is a non-monotonic function of mass. 
The higher is the mass, the higher is the resonance energy, $E_{\text{res}}$, given by Eq.~\eqref{eq:Eres2}.
This energy reaches $\mathcal{O}(100)$~MeV for $m_s \sim 30$~keV. 
For $E_{\text{res}}\gg T_{\max}$ the population of neutrinos is exponentially suppressed, switching the sterile neutrino production off as $m_s$ increases. 
For small masses of sterile neutrinos, they are copiously produced, but carry less energy ``per particle''.
As the mixing angle decreases for the fixed mass, the conversion probability~\eqref{eq:Px_s} decreases as well. As a result, the number of emitted sterile neutrinos drops, which explains why the contours close at small $\theta$.

At large mixing angles the situation is different. 
The resonance region increases with the increase of $\theta_0$ and eventually, becomes larger than the mean free path (c.f.~\eqref{kinetic_res}). This, again, destroys the resonance condition and conversion becomes non-efficient. This explains the upper boundary of the contours in Fig.~\ref{fig:EmittedEnergiesFullDiffusion}.

Formally, the maximal energy that can be carried by sterile neutrinos in our fiducial model is $E_s^{max} \approx 1.5\cdot 10^{53}$~erg, comparable with the total energy output in active neutrinos, $E_{\nu_\alpha} \simeq  \unit[10^{53}]{erg }$ (per flavor).
Such sterile neutrinos would be a significant cooling agent, affecting the temperature profile and effectively shutting down their own production. 
This back reaction has not been taken into account in our work and therefore their treatment is not done self-consistently. 
Therefore the red contours in Fig.~\ref{fig:EmittedEnergiesFullDiffusion} are definitely an overestimation and are shown only for the indication of the effect.
In order to properly account for sterile neutrinos with such a strong back-reaction, one would need a detailed numeric study. 
Here our goal was to demonstrate, that the back-reaction of lepton number production is still a significant effect. This comment is applicable also to other figures we present in the text.

\subsubsection{Quantifying the uncertainties}
\label{sec:uncertainties}

\begin{figure}[!t]
  \includegraphics[width=0.47\textwidth]{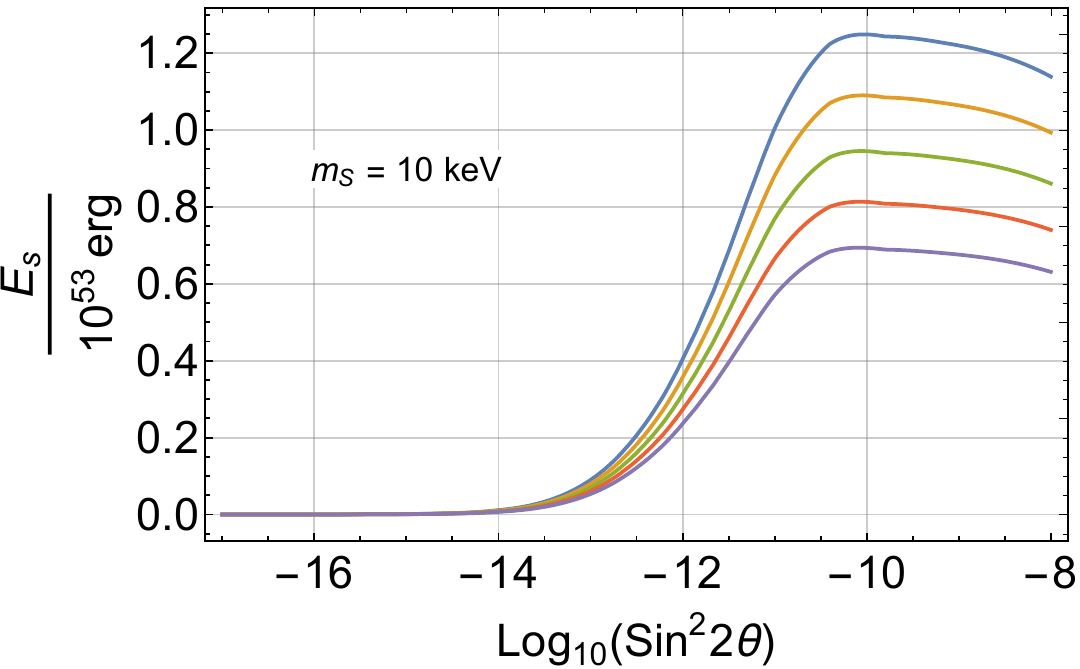} 
  \includegraphics[width=0.47\textwidth]{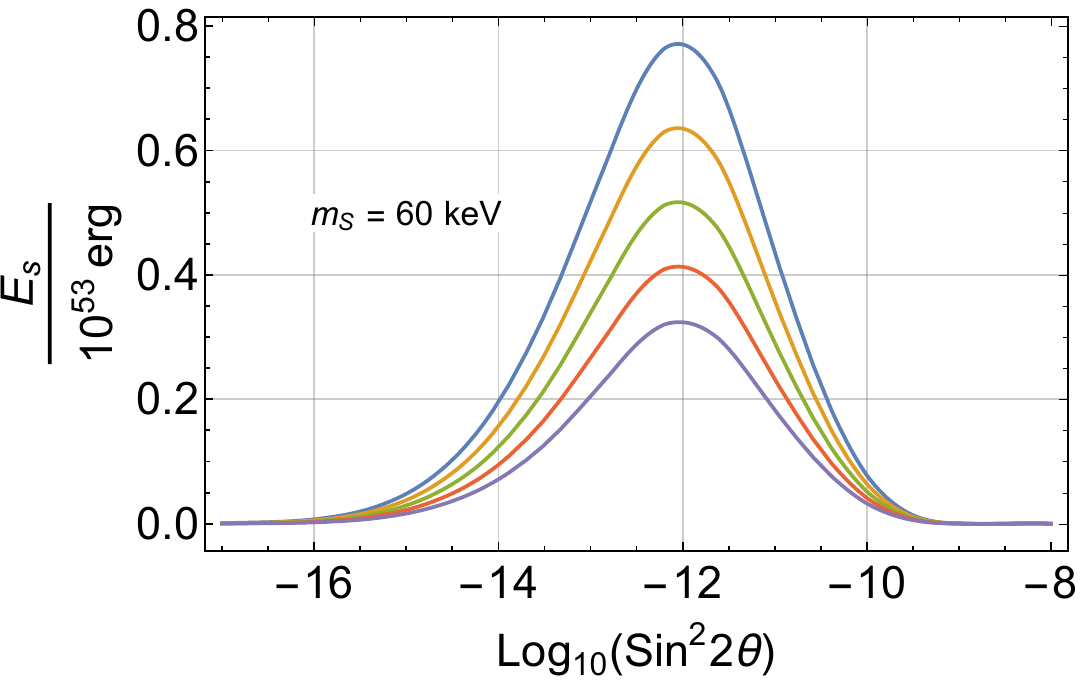}
  \caption{Energy, emitted in the form of sterile neutrinos for fixed masses $m_s = $ 10 keV (top plot) and 60 keV (bottom plot)  depending on the mixing angle. Different contours correspond to temperature value modifier starting from 1 (fiducial model) that produces the largest energy output, to the set of values 0.95, 0.9, 0.85, 0.8 as a sequence of contours with decreased production rate.
  We see that the model with $\sim 20\%$ smaller $T_{\text{max}}$ results in $3-4$-times lower energy yield in steriel neutrinos. }
  \label{fig:SN_model}
\end{figure}

 {\textcolor{black}The efficiency of the energy emission and, hence, our ability to set meaningful bounds on the sterile neutrino parameters is sensitive to the temperature in the post-bounce core. 
 This quantity is not known experimentally and can only be deduced from simulations.
Unfortunately, there is a range of viable models of supernova explosion
and they can provide quite different results regarding the parameters inside the supernova.
This is discussed in more detail in Appendix~\ref{app:SNmodel}.

Here, in order to indicate the level of uncertainties we repeat our calculations in the model with the same temperature profile,} suppressed in amplitude by $20\%$ -- a highly conservative estimate, as the uncertainty in temperatures can be much higher, see the comparison of temperature in two different simulations at Fig.~\ref{fig:Tmax_progenitor}. 
However, even these modifications can lead to significant changes in sterile
neutrino energy production.
Fig.~\ref{fig:SN_model} shows several additional ``slices'' at $m_s =
  \mathrm{const}$ that illustrates the dependence of our results on assumed inner
  temperature $T_{\text{max}}$.

\bigskip

\subsubsection{The importance of diffusion}
\label{sec:importance-diffusion}

\begin{figure}[!th]
  \includegraphics[width=0.45\textwidth]{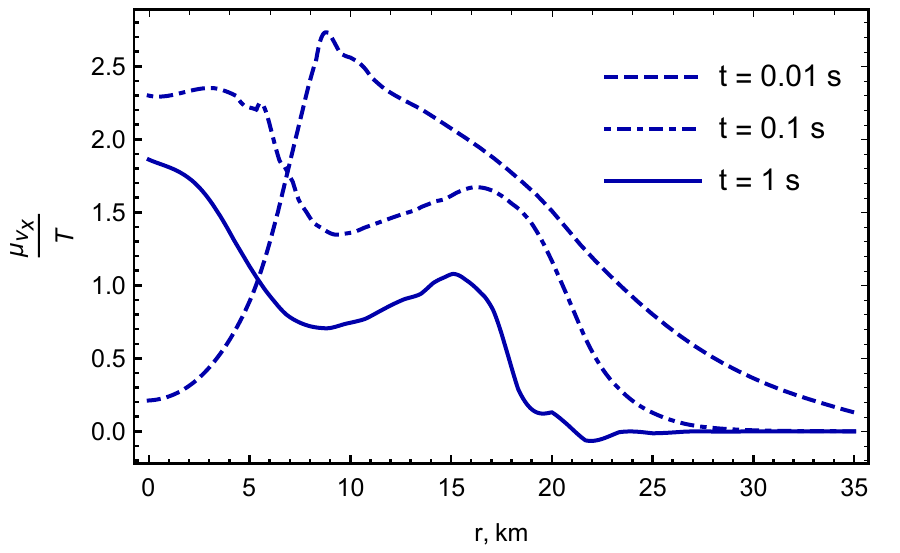} \hfill
  \includegraphics[width=0.45\textwidth]{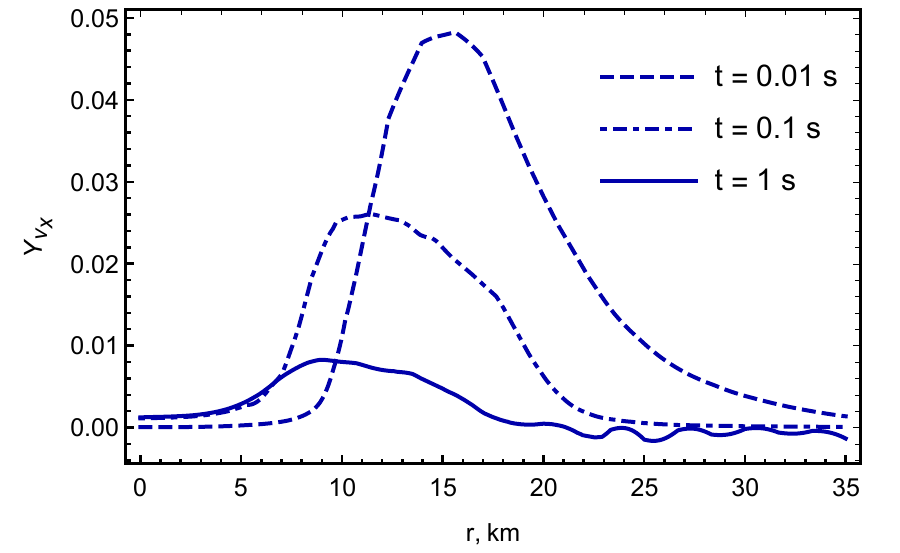}
  \caption{Time evolution of the radial profiles of the chemical potential
    $\mu_\tau$ and of the asymmetry parameter $Y_\tau$ in the fiducial model.
    Parameters of sterile neutrino are: mass $m_s = 7.1$~keV, the mixing angle
    $\sin^22\theta_\tau = 5\times 10^{-11}$.
    The production of asymmetry starts at radii $r = 10-20$~km, and then diffuses both to the inner region, where it remains partially trapped, and to the outer regions, where it can be carried away via neutrino emission.  
    Thus by $t\sim 1$~sec the chemical potential becomes negligible at $r\gtrsim20$~km while still being non-zero in the core region due to the rapid decrease of the density of the SN and, hence, the increase of the neutrino diffusion rate at larger radii.    }
  \label{fig:x-asymmetry_profile}
\end{figure}

The solution of Eq.~\eqref{diffsimplified_res} allows to find the evolution of the chemical potential $\mu_x$ that governs the distribution of active
anti-neutrinos. It is shown in Fig.~\ref{fig:x-asymmetry_profile}.
One sees that $\mu_x/T$ can reach significant values ($\mu_x \gtrsim T$).
\begin{figure}[!t]
  \centering \includegraphics[width=\linewidth]{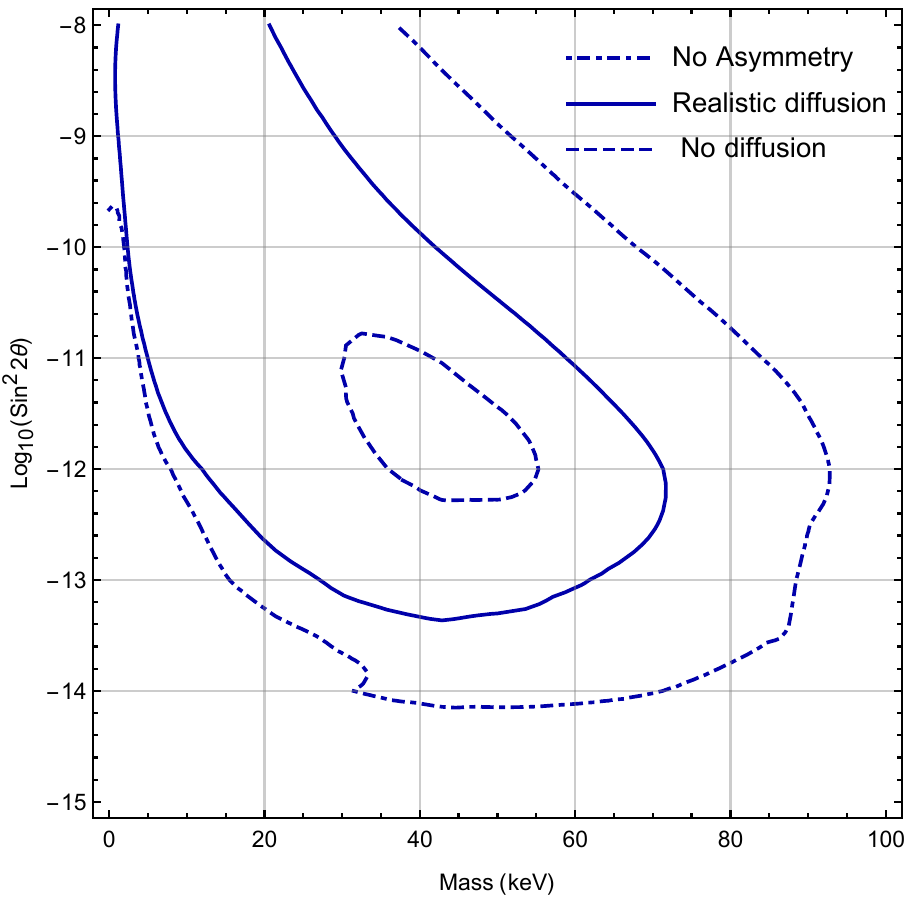}
  \caption{Effects of the  feedback. We show how energy contours ($E_s = 0.5 \cdot 10^{53} $erg) changes for three different feedback mechanisms: the depleted lepton number is not repopulated by any means (``\emph{no diffusion}'' dashed line); the restoration of the lepton number proceeds much faster than sterile neutrino production (``\emph{no asymmetry}'' dashed-dotted line); and the case of the realistic diffusion, as studied in this work.
    The mixing is with $\nu_{\tau}$ only and the duration of emission is taken to be 1~sec for all three cases.}

  \label{ThreeCases}
\end{figure}

To demonstrate the importance of back-reaction effects we also studied two extreme scenarios: \textit{(i)} the absence of diffusion and \textit{(ii)} the absence of back-reaction (infinite reservoir of neutrinos $\bar\nu_x$ at every energy and radius).
In the former case the production $\bar\nu_x\to \nu_s$ stops very quickly, as the resonant conversion ``consumes'' all active anti-neutrinos at a given radius and there are no mechanisms to replenish their population, as the large number of $\nu_x$ prevents the creation of $\nu_x\nux$ pairs via Pauli blocking.
(see also Appendix~\ref{asymmetryappendix} for more details).
Therefore the sizeable production of sterile neutrinos is possible in this case only for sufficiently large values of the mixing angle.
In the case \textit{(ii)}, the population of anti-neutrinos $\bar\nu_x$ gets immediately restored and therefore the conversion rate remains the same throughout the whole time $\tpb \sim 1$~sec, being extremely efficient.
The production in the case \textit{(ii)} stops only because neutrinos sufficiently cool down with the SN.
It is this approximation that was used in~\cite{Arguelles:2016uwb} which explains higher total energy emitted in sterile neutrinos in their case.
The realistic back-reaction is in-between these two limiting cases, as Fig.~\ref{ThreeCases} demonstrates.

\begin{figure}[!t]
  \centering \includegraphics[width=1\linewidth]{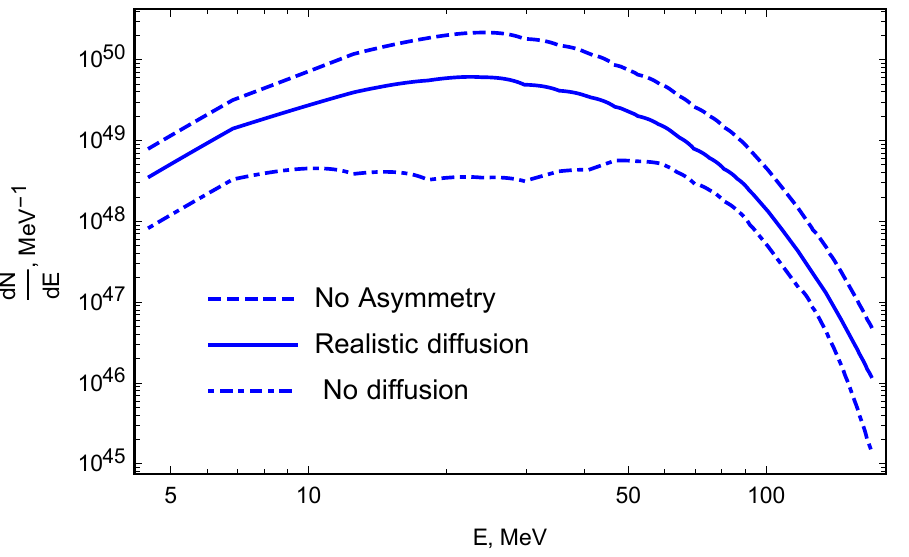}
  \caption{Spectra of sterile neutrinos with mass $m=7.1$~keV and the mixing
    angle $\sin^22\theta_x=5\times10^{-11}$ produced during the first second of
    explosion for three cases of different back reactions from Fig.~\protect\ref{ThreeCases}). Sterile neutrinos are mixed with $\tau$-flavor.
  }
  \label{Spectra}
\end{figure}

The spectra of the resulting sterile neutrinos with different diffusion treatment are shown in Fig.~\ref{Spectra}.

\subsubsection{Difference between muon and tau mixings}
\label{sec:mu-tau-diff}
Although the presented mechanism works for both $\mu$- and $\tau$-mixing, the treatment of these two flavors differs, due to the fact that the temperature of the SN interior, as well as  the value of the muon neutrino chemical potential $\mu_{\nu_{\mu}}$ (which appeared as a result of the back-reaction), is high enough for muon pairs to be present (but not for tau leptons):
\begin{equation}
  \label{eq:beta-eq}  
  \begin{aligned}
    Y_p ={} &Y_e + Y_{\mu}\\
  \mu_{e}-\mu_{\nu_e} ={} &\mu_n-\mu_p = \hat{\mu}\\
  \mu_{\mu}-\mu_{\nu_\mu} ={}& \mu_n-\mu_p = \hat{\mu}
\end{aligned}
\end{equation}
Once we fix SN-model dependent variables like total baryon density, the chemical potential of electrons, muon chemical potential can be calculated which will affect the neutrinos effective potential (Eq.~\eqref{Veff}). In the case of $\tau$-leptons, their mass is too high even with non-zero $\mu_{\nu_{\tau}}$ to be produced. But even in the case of muon neutrinos production, the achieved $Y_{\mu} << 0.1$ hence does not affect the production at a noticeable level compared to tau-flavor mixing and there is no difference in the resulting amount of energy, carried by either flavor.\footnote{As the change in the effective potential and hence the resonance energy was too small.}
Therefore, our results (Fig. \ref{fig:EmittedEnergiesFullDiffusion}) does not depend on mixing flavor. We do not discuss here the influence of charged muons on the SN explosion~\cite{Bollig:2017lki}.

\subsubsection{Damping}
\label{sec:damping}

The neutrino damping~\cite{Stodolskiy1987} describes the probability that a neutrino would interact with the medium while propagating in the resonance region.
This interaction will cause the wave function to collapse to a pure flavor state, and its resonance conversion will become impossible.
One can ignore the damping whenever $R_\width \ll \lambda_\mfp$.
In the opposite limit, the collisional production becomes important,  as the scattering has a finite probability to leave behind not only pure active but also pure sterile state.  
The collisional production has been considered before in many works (see \emph{e.g.}~\cite{Abazajian:2001nj,Raffelt:2011nc,Arguelles:2016uwb,Suliga:2019bsq}) and it is beyond the scope of the current work to study how it combines with the resonant production.

 Its effect can be understood as follows: the width of the resonance $R_\width$ is independent on the energy and proportional to the $\sin (2 \theta_0)$ (Eq.~\eqref{eq:Rwidth}) while the mean free path of active neutrinos scales with energy as $E^{-2}$ (see Section~\ref{app:diffusion}). 
As a result for a given mass $m_s$ and position $R_\res$ (equivalently \textit{fixed resonance energy}) the ratio $R_\width/\lambda_\mfp$ grows with~$\theta_0$.
If one keeps the mixing angle (and $R_\width$) fixed, but rather increases the mass -- the resonance energy is increasing (Eq.~\eqref{ResonanceEnergy_res}). Therefore the mean free path of $\nux$ decreases and neutrino damping becomes important.

\subsubsection{Sterile neutrino as dark matter}
\label{sec:sterile-neutrino-DM}

So far we did not make any reference to sterile neutrinos being dark matter particles.
The lifetime of sterile neutrinos lighter than two electron is given
by (assuming for simplicity that $\theta_x$ is the only non-negligible mixing)
\begin{equation}
  \label{eq:11}
  \tau_s \approx \unit[2\times 10^{24}]{sec}\parfrac{10^{-11}}{\sin^2(2\theta_x)}\parfrac{\unit[20]{keV}}{m_s}^5
\end{equation}
-- much longer than the lifetime of the Universe when $\theta^2 \sim 10^{-11}$.
And indeed such particles represent a viable dark matter candidate
(as suggested in \cite{Dodelson:1993je,Shi:1998km,Dolgov:2000ew,Abazajian:2001nj,Asaka:2006nq}, see~\cite{Boyarsky:2018tvu} for a review).

We compute the energy output for a sterile neutrino with mass $m_s= 7.1$~keV and mixing angle $\sin^2{2\theta_x} = (2-20) \times 10^{-11}$.
Decay of such a sterile neutrino dark matter would produce an X-ray line, consistent with the observations of~\cite{Bulbul:2014sua,Boyarsky:2014jta} and many subsequent works, see~\cite{Boyarsky:2018tvu} for details.
In this case, the energy output would be $E_s \eqsim 1.5\cdot 10^{53}$ erg.

The grey shaded region in Fig.~\ref{fig:EmittedEnergiesFullDiffusion} shows the parameter space of the \emph{Neutrino Minimal Standard Model} ($\nu$MSM)~\cite{Asaka:2005pn,Asaka:2005an}, see~\cite{Boyarsky:2009ix} for review where sterile neutrinos would have correct dark matter abundance (parts of this parameter space are excluded by X-ray and structure formation constraints, see~\cite{Boyarsky:2018tvu} for review.
The upper boundary corresponds to the parameters of the non-resonant dark matter production~\cite{Dodelson:1993je,Abazajian:2001nj,Asaka:2006nq}, while in the rest of the region the correct dark matter abundance can be obtained in the presence of primordial lepton asymmetry~\cite{Shi:1998km,Abazajian:2001nj,Shaposhnikov:2008pf}.
The maximal value of lepton asymmetry required to produce the correct dark matter abundance depends on the ratio of the mixing angles and differs, for example, in the model where $\theta_e = \theta_\mu = \theta_\tau$ as opposed to that with only $\theta_\tau \neq 0$~\cite{Shaposhnikov:2008pf,Ghiglieri:2015jua}.
We conservatively chose to plot the lower bound corresponding to the maximal value of the lepton asymmetry attainable in the $\nu$MSM~[\citenum{Shaposhnikov:2008pf},\,\citenum{Boyarsky:2009ix}].

\section{Discussion}
\label{sec:conclusion}

In this paper, we analyzed the process of sterile neutrino creation during the explosion of a core-collapse supernova.
Sterile neutrinos are produced via mixing with active anti-neutrinos of $\mu$ and/or $\tau$ flavors (collectively, $\bar\nu_x$).
The hot and dense supernova environment is non-transparent for neutrinos and their dispersion changes as compared to the propagation in a vacuum.
Therefore, the mixing with sterile neutrinos can become \textit{resonant} (the MSW-like effect), leading to the effective conversion of anti-neutrinos $\bar\nu_x$ into sterile neutrinos with mass in the range $\unit[5]{keV} \lesssim m_s \lesssim \unit[40]{keV}$ and mixing angles $\sin^2(2\theta_x)$ reaching $10^{-8}$ and below.
The question of sterile neutrino production during supernovae explosion, their effects on explosion, and on the stellar nucleosynthesis has been studied in the past for sterile neutrinos ranging in masses from eV to GeV~\cite{Kainulainen:1990bn,Raffelt:1992bs,Peltoniemi1992,Shi:1993ee,Nunokawa:1997ct,Kusenko:1997sp,Dolgov:2000pj,Dolgov:2000jw,Dolgov:2000ew,Abazajian:2001nj,Fuller:2003gy,Barkovich:2004jp,Hidaka:2006sg,Hidaka:2007se,Kusenko:2008gh,Fuller:2009zz,Tamborra:2011is,Raffelt:2011nc,Wu:2013gxa,Warren:2014qza,Zhou:2015jha,Warren:2016slz,Arguelles:2016uwb,Rembiasz:2018lok,Xiong:2019nvw}.
With few exceptions (\textit{e.g.}\ \cite{Abazajian:2001nj,Raffelt:2011nc,Zhou:2015jha,Arguelles:2016uwb}) these studies concentrated on the mixings of sterile neutrino with electron flavor.
Recent work~\cite{Arguelles:2016uwb} argued that the fast production of sterile neutrinos is possible due to the MSW-like resonance outside the SN core region when mixing with \nux.
However, the authors of~\cite{Arguelles:2016uwb} did not account for the depletion of the population of $\bar\nu_x$ in the resonance region and kept the distribution of active anti-neutrinos at its equilibrium level, thus providing a ``stock'' of anti-neutrinos to be converted.
In reality, the depletion of the active anti-neutrinos slows down the conversion process; the $\nu_x-\bar\nu_x$ pair creation re-populates the abandoned states, and the above-equilibrium excess of $\nu_x$ gets diffused away.

In this work, we properly took into account the diffusion of the lepton number and the back-reaction of sterile neutrinos on the neutrino distribution.
Our results show that sterile neutrinos can carry away the amount of energy, comparable to that of active neutrino flavors (see Fig.~\ref{fig:EmittedEnergiesFullDiffusion}).
While the energy output can reach $10^{53}$~ergs -- a ballpark figure associated with an SN explosion -- \emph{this does not lead to the bounds that are both strong and robust.}

Indeed, two main types of bounds from supernovae exist: \emph{energy loss} and \emph{energy-loss rate} bounds, see \emph{e.g.}~\cite{Raffelt:1990yz,Raffelt:1996wa,Raffelt:1999tx,Adhikari:2016bei}.
The emission of any exotic component can be capped from above by $E_\tot$ -- the total energy available in an explosion.
The latter is the difference between the binding energies of a progenitor and a remnant.
The estimates of the total released energy $E_\tot$ depend on whether the remnant is a black hole or a neutron star.
It is generally believed that the remnant of SN1987A is a neutron star, although the remnant has not been found~\cite{Alp:2018oek} after more than 30 years of searches.
The NS remnant can still be hidden behind SN debris~\cite{Esposito:2018nib,Alp:2018oek} and there is a rising possibility that the remnant is indeed the NS according to recent work \citenum{Page:2020gsx}.
If the remnant is the neutron star, its binding energy can be estimated as
\begin{equation}
  \label{eq:grav_binding}
  E_{\rm NS} 
  \approx \unit[6.3 \times 10^{53}]{erg}\parfrac{\mathcal{C}}{0.6}\parfrac{M_{\rm NS}}{2 M_\odot}^2 \parfrac{10\km}{R_{\rm NS}}
\end{equation}
with the coefficient $\mathcal{C} \approx 0.6$~\cite{Lattimer:2000nx,Lattimer:2005aj,Janka:2017vlw}.
The estimates put the mass for the SN1987A remnant in the range $M_{\rm NS} \simeq 1.7- 1.9 M_\odot$, see~\cite{Alp:2018oek} for review.
Alternative scenarios for a black hole formation in the SN1987A explosion  exist~\cite{1992ComAp..16..153B,Beacom:2000qy,Blum:2016afe,Bar:2019ifz}.
In any case, the energy emitted in sterile neutrinos (Fig.~\ref{fig:EmittedEnergiesFullDiffusion}) is smaller than $E_{\rm NS}$.

\textcolor{black}{The energy loss rate argument} \cite{Raffelt:1990yz, Raffelt:1996wa,Raffelt:1999tx} \textcolor{black}{$ \epsilon_{\rm extra} \lesssim \unit[10^{53}]{erg/sec} $ is based on the shortening of the active neutrino signal duration in presence of additional cooling channel.The corresponding study was provided for the case of axions} ~\cite{Burrows:1988ah,Keil:1996ju,Fischer:2016cyd}\textcolor{black}{  and although there might be differences in details of production mechanisms (namely, the area of production in the case of the resonant neutrino production correspond mostly to regions, that are located outside the core and up to neutrinosphere while axions are produced the most intensively in the core), we can expect the same order-of-magnitude constraint.} The same bound, of course, can be applied for sterile neutrinos, produce via scatterings~\cite{Dolgov:2000pj,Abazajian:2001nj,Fuller:2003gy,Fuller:2009zz,Raffelt:2011nc}).

In addition to the previous points, the output of sterile neutrinos is
sensitive to the temperature (and temperature profile) in the inner regions of the SN.
For $m_s \sim \mathcal{O}(\SI{1}{keV})$  the available neutrino population scales as $E_{\res}^{2}$ in the whole SN region where the condition $E_{\res}(R) \ll T$ holds. 
\textcolor{black}{For $m_s \sim \mathcal{O}(\SI{100}{keV})$, since $E_{\res} \gg T$ everywhere, the number of ``available'' neutrinos scales exponentially with the inner temperature.
The temperature dependence is thus more pronounced for the higher mass sterile neutrinos.}

No observables are sensitive to the temperatures in these regions as the emission of active neutrinos happens from the outer regions -- the neutrinosphere with $R_{\nu\sph} > R_\res$.
Therefore, even detailed measurements of the neutrino fluxes would not tell us about the conditions under which sterile neutrinos were produced.
Knowledge of the temperature profile (that would allow recovering
$T(R_\res)$ given the ``measurement'' of $T(R_{\nu\sph}$) can only be inferred
from the simulations  (similar
to \textit{e.g.}~\cite{Warren:2014qza,Rembiasz:2018lok} that however deal with
heavier sterile neutrinos and/or different production mechanisms and influence
on the SN dynamics).
Such bonds will necessarily be model-dependent.
We leave the self-consistent treatment of these cases to future works.

Finally, we note that the same challenges are faced by energy loss bounds applied to other hypothetical very weakly interacting particles: axions, dark photons, millicharged particles, etc. 

\section*{Acknowledgements}

We would like to thank G.~Fuller, G.~Raffelt, I.~Tamborra, I.~Timiryasov, and Y.-Z.~Qian for many useful discussions and comments on the draft.
O.R.\ thanks the organizers of the \textit{``Neutrino Quantum Kinetics in Dense Environments''} workshop for creating a stimulating environment.
This work was supported by the Carlsberg Foundation and by the European Research Council (ERC) under the European Union’s Horizon 2020 research and innovation program (GA 694896).

\appendix
\numberwithin{figure}{section}
\numberwithin{equation}{section}

\onecolumngrid
\section{The fiducial supernova model}
\label{app:SNmodel}

\begin{figure}[!t]
\includegraphics[width=0.5\textwidth]{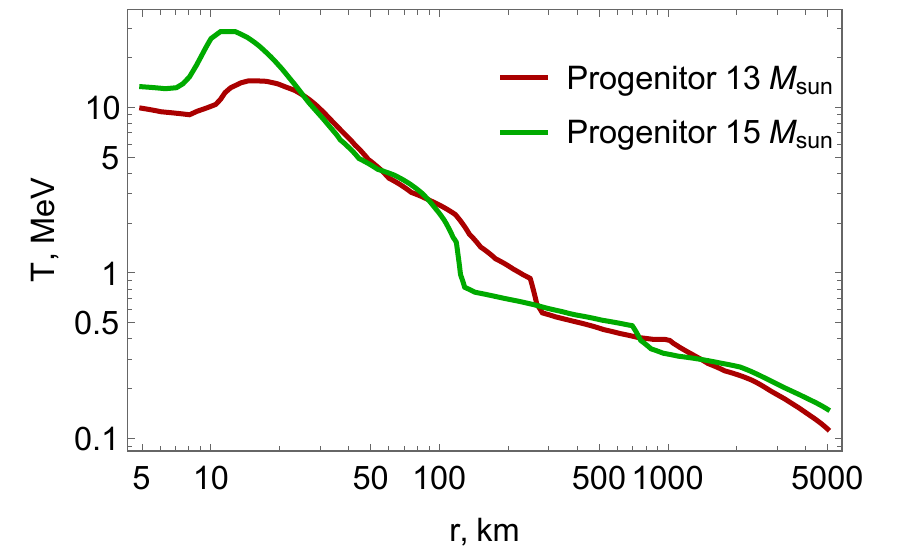}
\caption{Dependence of the temperature profiles (and in particular of the maximal temperature, $T_{\max}$) on the mass of progenitors. Both temperature profiles are for the same post-bounce time $t_{pb} \simeq \unit[250]{msec}$ and obtained as a result of simulations with the same numerical code \cite{Liebendoerfer:2003es}.
The plots are shown for two different progenitor models with the main sequence
masses of $13M_\odot$~\cite{Nomoto:88} and $15M_\odot$~\cite{Woosley:95} that provided the initial
conditions for the corresponding runs.
The uncertainty in the determination of the mass of the progenitor of the SN1987A is $15-20M_\odot$~\cite[see e.g.][]{Weiss:89}. \textcolor{black}{Although the selected time for a snapshot does not correspond to the period, when everything in the SN is settled down, we believe it perfectly demonstrates mentioned potential discrepancy in the SN media conditions during the explosion which can be relevant for the sterile neutrino production}
 }
\label{fig:Tmax_progenitor}
\end{figure}
The main goal of our paper is to demonstrate the effect of back-reaction from the build-up of the lepton asymmetry on the resonant production of sterile neutrinos.
The sterile neutrino emission depends on the spatial and temporal distribution of density of baryons $\rho_B$, temperature, asymmetries of electrons and of neutrinos $Y_e$, $Y_{\nu_\alpha}$. 
These quantities cannot be measured directly and in general require the numerical solution of a system of hydrodynamic transport equations to learn something about their properties. This introduces a number of systematic uncertainties.

Different numerical approaches to the supernova give broadly consistent results (see e.g.\ the comparison of codes and approximations in \cite{Just:2018djz,OConnor:2018sti}).
Typical differences in various observables obtained with different codes are $\mathcal{O}(10\%)$.
On the other hand, different assumptions about the SN  progenitors can lead to very different temperature profiles (under otherwise equivalent assumptions), see example in Figure~\ref{fig:Tmax_progenitor}.\footnote{The progenitor of the SN1987A is a blue supergiant star Sk~$-62^\circ 202$ \cite{Gilmozzi:87} whose mass is estimated to be in the range $15-20M_\odot$~\cite{Weiss:89}.}
The difference of $T_{max}$ can lead to order-of-magnitude changes in the number of produced sterile neutrinos (see Appendix~\ref{app:uncertainties}) at fixed mass and mixing angle.

{\textcolor{black}Another important uncertainty comes from the unknown equation of state (EoS) of nuclear matter. 
Different equations of state (see \textit{e.g.} \cite{Shen:1998gq,Shen:2011kr,Lattimer:1991nc,Hempel:2009mc}) appear as a result of different treatment of nuclear matter and its composition, see \textit{e.g.} \cite{Pons:2000iy}.
The evolution of proto-neutron stars and corresponding neutrino signal under the assumption of different EoS were actively studied \cite{Reddy:1998gc, Prakash:2000jr,Pons:1998mm,Prakash:1996xs,Hempel:2017ikt,Steiner:2012rk,Reddy:1997yr,Sumiyoshi:2006id,Sumiyoshi:2008zw}. 
The nuclear equation of state can even be decisive in whether the simulation of an explosion would be successful~\cite{Sumiyoshi:2006id,Sumiyoshi:2008zw,OConnor:2010moj,Fischer:2020xjl}. 
Overall, depending on the nuclear equation of state, the parameters that are crucial for the production of sterile neutrinos -- temperature, density, and lepton asymmetries -- can vary significantly (see, for example, comparison of numeric results in \cite{Prakash:1996xs,Steiner:2012rk}).}

Given all these uncertainties, in this work, we purposely do not establish any constraints and demonstrate that the current state of the art (both observational and theoretical) does not allow us to provide any robust constraints.

\begin{figure}[!h]
  \includegraphics[width=0.47\textwidth]{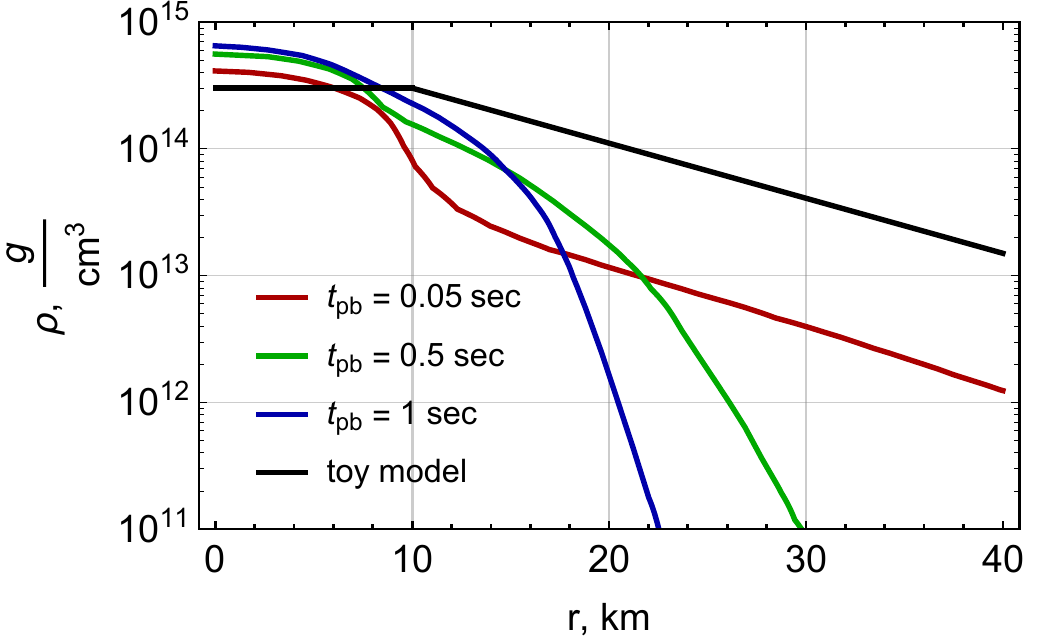} 
  \includegraphics[width=0.47\textwidth]{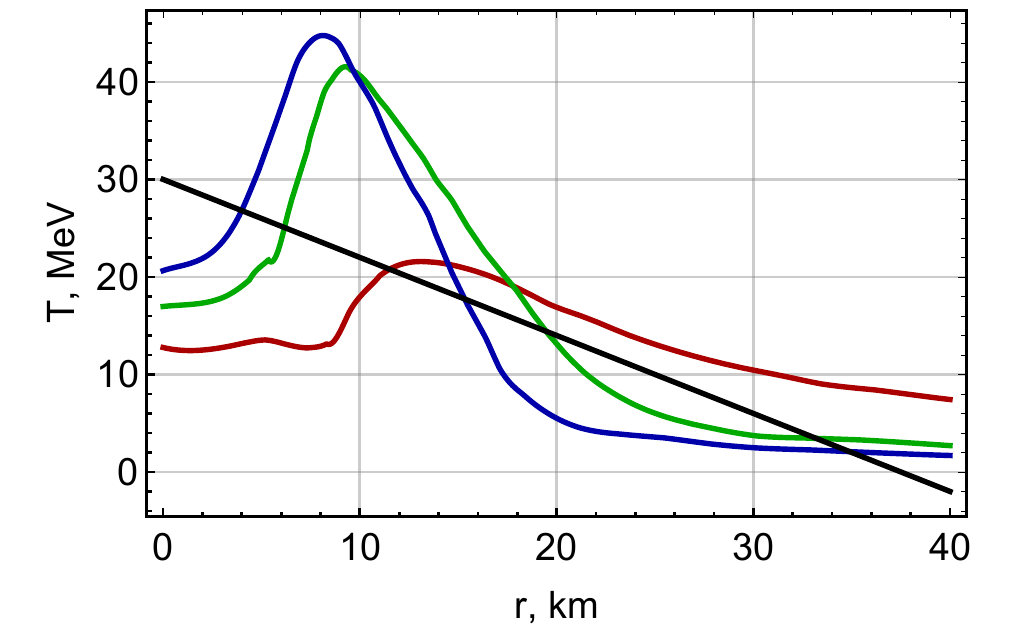}\\
  \includegraphics[width=0.47\textwidth]{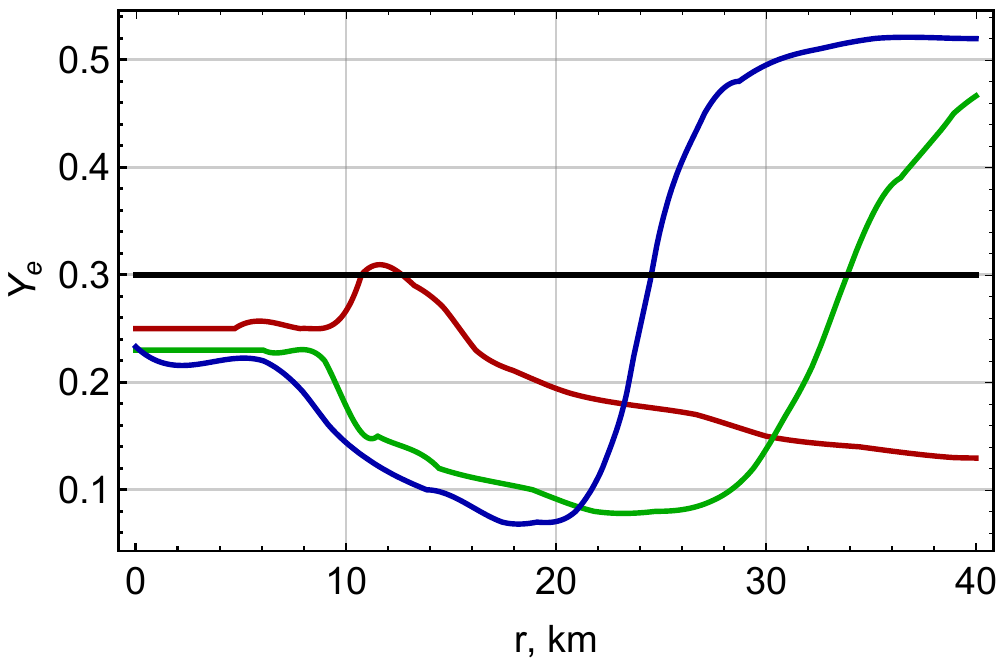} 
  \includegraphics[width=0.47\textwidth]{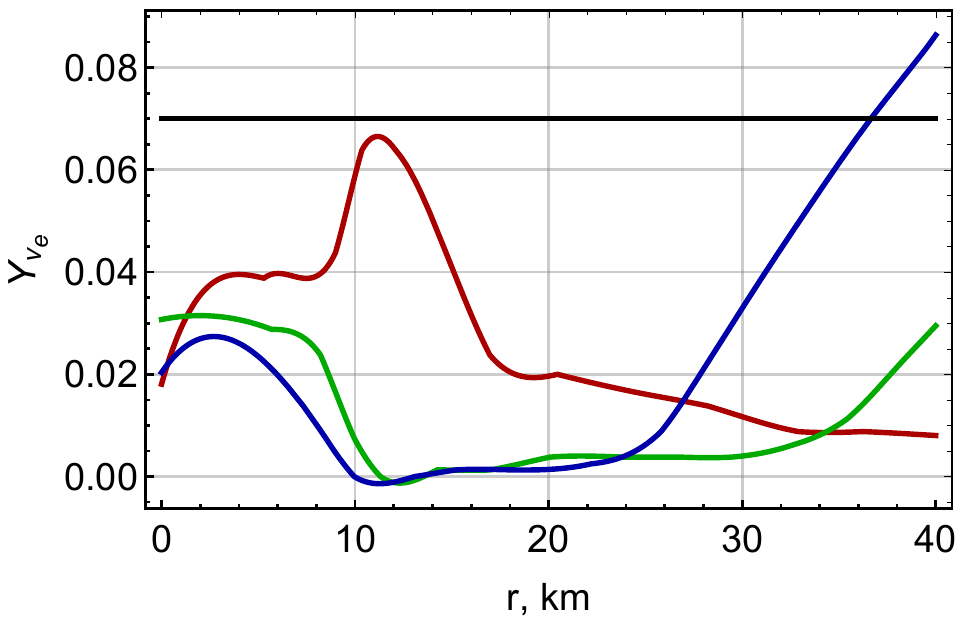}
  \caption[Radial profiles of relevant quantities in the fiducial model.]{Radial profiles of density, temperature, electron and electron neutrino asymmetries, taken as snapshots  from 1D hydrodynamic simulations of the 18.6 $M_{\odot}$ supernovae explosion~\cite{Garchinv_archive}. Post-bounce times are $t_{pb} = 0.05, 0.5, 1$~sec. Black lines show the (time-independent) profiles in our toy model, used in Appendix~\ref{app:uncertainties} below.
  }
  \label{fig:SN_model_app}
\end{figure}

However, in order to perform the analysis and estimates the magnitude of the described effects, we adopt a \emph{fiducial  SN model}, compute sterile neutrino production within it, and then quantify possible uncertainties. 
Our model is based on a 1D hydrodynamic simulation of an SN model~\cite{Garchinv_archive} with the progenitor mass of $18.6 M_\odot$ and SFHo nuclear equation of state~\cite{Steiner:2012rk} and the gravitational mass of $1.4M_\odot$.
\textcolor{black}{To allow for simplified analytical treatment of the problem, instead of using the exact temporal evolution of the SN background we use a model, when we have three snapshots for density, temperature and electron asymmetry profile obtained in simulation  at post-bound times $t_{\text{pb}} = 0.05, 0.5, 1$~sec (see Fig.~\ref{fig:SN_model_app}).
We use these parameters from snapshots as static background during the correspondent time intervals ( $0 \leq t < 0.05 $, $0.05 \leq t < 0.5 $,  $0.5 \leq t < 1 $ ) and evolve the HNL production as well as $\mu/\tau$-asymmetry over this static background. So, for every new time interval, the initial profile of the lepton asymmetry is taken from the previous step evolution. While keeping the calculation as simple as for the completely static profile, this allows to follow the changes in production rate during different post-bounce times.}

Somewhat similar model and a similar approach have been recently used e.g.\ in \cite{Suliga:2019bsq,Suliga:2020vpz}.
At times $\tpb > 1$~sec, the temperature drops down to the values below few~MeV, which
results in a low rate of $\nu_s$ creation.  
That is why we do not take into account times $t > \unit[1]{sec}$.

\section{Resonant sterile neutrinos production}
\label{ResConversion}

For completeness, we reproduce the formalism of the resonant conversion for neutrinos propagating in the media with changing density. 
Each of the flavor states $\nu_x,\bar\nu_x$ as well as $\nu_s$ obeys the Dirac
equation and as a consequence the Klein-Gordon equation.  When particles are
ultra-relativistic in the medium of variable density this equation can be
brought into the form (see \textit{e.g.}\ the book~\cite[Chap.~8]{Raffelt:1996wa}):
\begin{equation}
  \label{eq:10}
  i \frac{d}{dr} 
  \begin{pmatrix}
    \bar\nu_x                                                            \\
    \nu_s
  \end{pmatrix} = \mathcal{H}_\eff(r) \begin{pmatrix}
    \bar\nu_x                                                            \\
    \nu_s
  \end{pmatrix}
\end{equation}
where the ``effective Hamiltonian'' is
\begin{equation}
  \label{Heff}
  \mathcal{H}_{\rm eff}(r) = \frac{m_s^2}{4E}
  \begin{pmatrix}
    -\cos 2\theta_0                                                      & \sin2\theta_0                                        \\
    \hphantom{-}\sin2\theta_0 & \cos2\theta_0
  \end{pmatrix}+
  \begin{pmatrix}
    V_{\rm eff}(r)                                                       & 0                                                    \\
    0 & 0
  \end{pmatrix}.
\end{equation}
Here  $V_\eff$ is the effective potential of \nux given by (see Eq.~\eqref{Veff} for details/notations):
\begin{equation}
      V_{\eff}(r) = -\frac{G_F}{\sqrt{2}}N_{\rm b}\Bigl(Y_n - 2Y_{\nu_e} - 2Y_{\nu_\tau} - 4Y_{\nu_\mu}-2Y_{\mu}\Bigr) = \unit[11.4]{eV}\parfrac{N_{\rm b}}{N_0}\Bigl(Y_n - 2Y_{\nu_e} - 2Y_{\nu_\tau} - 4Y_{\nu_\mu}-2Y_{\mu}\Bigr)\;,
\end{equation}
$m_s$ is the mass of sterile neutrino, $E$ is its energy ($m_s \ll E$)
and we have neglected masses of the active neutrinos; $\theta_0$ is the vacuum
active-sterile mixing angle. 
The sign of $V_\eff$ is such that only the mixing $\bar\nu_x - \nu_s$ is relevant and therefore we have omitted $\nu_x$ state in Eq.~\eqref{eq:10}.

{For future convenience we will introduce the notation
  \begin{equation}
    \label{eq:7}
    \Delta_s = \frac{m_s^2}{2E}
  \end{equation}
  When $V_\eff = 0$ the eigenvalues of the Hamiltonian~\eqref{Heff} are
  $\pm \frac 12 \Delta_s$ and the vacuum active-sterile oscillation length is
  given by $\pi/\Delta_s$.}

{ Notice that $[\mathcal{H}_\eff(r), \mathcal{H}_\eff(r')] \neq 0$ for
  $\theta_0 \neq 0$ and therefore exact solution of Eq.~\eqref{eq:10} is
  complicated.  For the propagation inside the star where
  $|\nabla \log V_\eff| \ll \Delta_s $ one can, however, solve this equation
  in the adiabatic limit.  To this end one diagonalizes~\eqref{Heff} at every
  point by the matrix $U(r)$, given by
  \begin{equation}
    \label{eq:8}
    U(r) =
    \begin{pmatrix}
      \hphantom{-}\cos\theta(r) & \sin\theta(r) \\
      -\sin\theta(r)            & \cos\theta(r)
    \end{pmatrix}
  \end{equation}
  where the \emph{matter mixing angle} $\theta(r)$ is defined (assuming
  $\theta_0 \ll 1$)}
\begin{equation}
  \label{eq:mixing_angle}
  \tan 2\theta(r) \simeq 2\theta_0\frac{\Delta_s}{\Delta_s + V_{\rm eff}(r)} + \mathcal{O}(\theta_0^2)
\end{equation}
From Eq.~\eqref{eq:mixing_angle} one sees that deep inside the SN, where $\Delta_s < |V_\eff(r_{\rm in})|$ and $V_\eff < 0$, one has $\tan 2\theta_{\rm in} \to -0 \Leftrightarrow \theta_{\rm in} \to \frac\pi 2$, because $\theta$ is confined to $0 \le \theta \le \frac\pi 2$.
On the other hand, when the condition
\begin{equation}
  \Delta_s + V_{\rm eff}(r) = 0
  \label{rescondition}
\end{equation}
is satisfied, one has a resonance and $\theta_\res \rightarrow \frac{\pi}{4}$.  Due
to the sign of effective potential, resonance condition \eqref{rescondition}
can be satisfied only for anti-neutrinos.  { Eq.~\eqref{rescondition}
  establishes a relation between the anti-neutrino energy and the radius of
  the resonance, $R_\res$:
  \begin{equation}
    V_\eff(R_\res)= -\frac{m_s^2}{2 E }
    \label{eq:Eres}
  \end{equation}}
which leads to Eq.~\eqref{ResonanceEnergy_res}.

{Diagonalisation of the Hamiltonian~\eqref{Heff} gives two eigenvalues
  $E_{a,b}(r)$ such that
  \begin{equation}
    \label{eq:9}
    E_{a,b}(r) = \frac{V_\eff}2 \pm \sqrt{(\Delta_s + V_\eff)^2 + 4  \Delta_s^2 \theta_0^2}
  \end{equation}
  and two eigenfunctions (mass eigenstates) $\nu_{a,b}$.}  In the medium with
variable density the states $\nu_{a,b}$ propagate according to the equation,
similar to Eq.~\eqref{eq:10}:
\begin{eqnarray}
  \label{evolution}
  i \frac{d}{dr}\left( 
  \begin{array}{c} 
    \nu_a                                                                \\
    \nu_b
  \end{array} \right)                                                    &=                                                     &  
                                                                                                                                  \begin{pmatrix}
                                                                                                                                    E_a(r)                                       & i {\theta'}(r)                                       \\
                                                                                                                                    -i
                                                                                                                                    {\theta'}(r)
                                                                                                                                    &
                                                                                                                                    E_b(r)
                                                                                                                                  \end{pmatrix}
                                                                                                                                      \begin{pmatrix}
                                                                                                                                        \nu_a                    \\
                                                                                                                                        \nu_b
                                                                                                                                      \end{pmatrix}
\end{eqnarray}
The off-diagonal elements in the r.h.s. are equal to
$-i U^\dagger \partial_r U$ and are responsible for transition between
different mass eigenstates that would be absent for $\theta'= 0$.  Let us
introduce a parameter of non-adiabaticity, $\gamma$
\begin{equation}
  \label{eq:gamma}
  \gamma \equiv \frac{\theta'(r)}{E_a(r) -E_b(r)}
\end{equation}
\begin{figure}[!t]
  \includegraphics[width=0.5\textwidth]{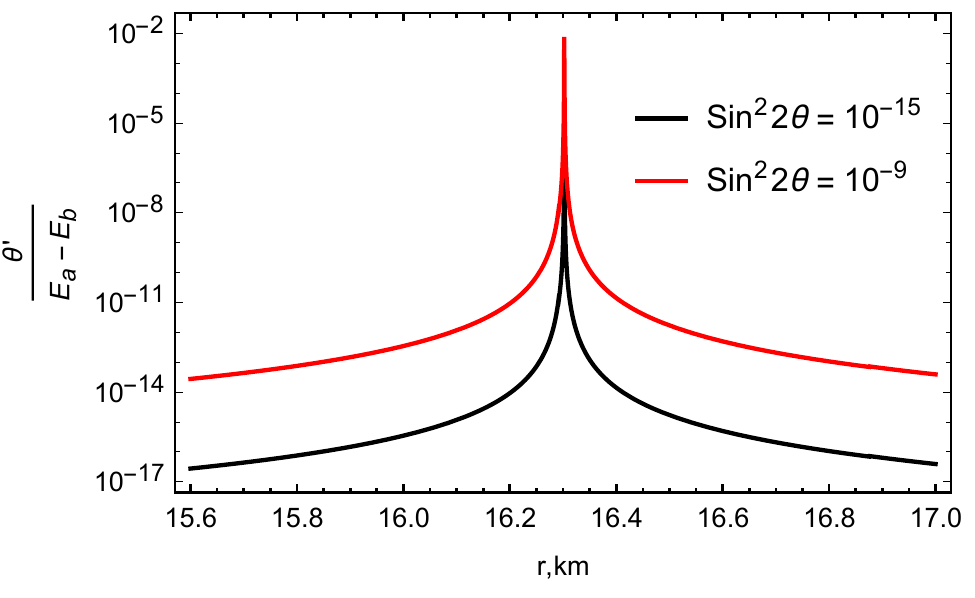}
  \caption{Evolution with radius of the adiabaticity  parameter $\gamma = \frac{\theta'}{E_a -E_b}$ (Eq.~\eqref{eq:gamma}).   
    It can reach large values in a very narrow region around the resonance ($R_\res \simeq \unit[16.3]{km}$ in this case) and is extremely small outside it.
    The energy of the neutrino equals $E=40$~MeV, sterile neutrino mass $m_s = 10$~keV. For smaller angle, the value of this parameter can be larger than 1. It shows that conversion goes non-adiabatically while for larger angle it acquire value $\ll 1$ everywhere, so the conversion is totally adiabatic}.
  \label{fig:gamma}
\end{figure}
\begin{figure}[!t]
  \centering \includegraphics[width=0.45\textwidth]{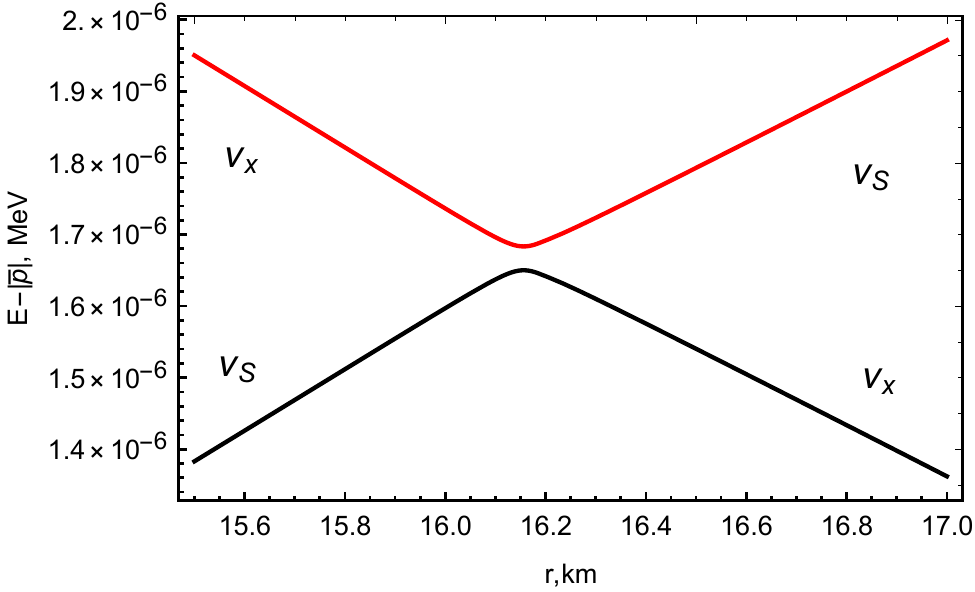}
  \caption{Energy levels $E_a$({\textcolor{black} black}), $ E_b$ (red) of the system, depending on the radius.  Mixing angle is chosen as
    $\sin^22\theta = 10^{-3}$, mass $m_s = 10$~keV, momentum $p = 30$~MeV. The $y$-axis shows $m_{a,b}/{2E}$. The closest distance between energy levels is at the resonance where the transition between the levels is the most likely.}
  \label{EnergyLevels}
\end{figure}

Its value determines whether a transition between different levels is
possible.  When $\gamma \to 0$, the evolution is fully adiabatic and
transitions between mass eigenstates are negligible (this is the case, for
example, in the Sun).  It turns out that for small $\theta_0$ $\gamma$ can be
different from zero only in a narrow region around the resonance for a wide
range of densities (effective potential) profiles (see Fig.~\ref{fig:gamma}).
This is the region where the mixing angle changes its value significantly.
Defining this region as where $\theta(r)$ changes from $\sin^22\theta = 1$ to $\sin^22\theta =\frac{1}{2}$ (\textit{i.e.} $\frac\pi8 \le \theta(r) \le \frac{3\pi}8$) we  get its width $R_{\width} = 2\sin(2\theta_0)/ \bigl(\log V_\eff(R_\res)\bigr)'$,  Eq.~\eqref{eq:Rwidth}.
The non-adiabaticity parameter is maximal at the resonance and can be expressed through the width of the resonance
  \begin{equation}
    \label{eq:gamma2}
    \gamma     = \frac{1}{\pi}\frac{L_{\osc}}{R_{\width}}
\end{equation}
where $L_{\osc}$ is the oscillation length at resonance~\eqref{eq:Losc}.
The probability of transition between mass states $\nu_a$ and $\nu_b$ after
crossing the resonance is given by \cite{Parke:1986jy}
\begin{equation}
  P_{x\to s} =  \frac{1}{2}- \left(\frac{1}{2}-P_{\rm na}\right) \cos2\theta_{\rm in} \cos2\theta_{\rm out}
  \label{presfull}
\end{equation} 
where $\theta_{\rm in} \simeq \frac\pi 2$ -- mixing angle, at the point of neutrino state creation and $\theta_{\rm out} = \theta(r_{\rm out}) \simeq \theta_0$ --  the vacuum mixing angle.
$P_{\rm na}$ is a probability of transition between mass eigenstates due to non-adiabatic change of $V_{\eff}$.
In the case, when $R_{\width}$ is much smaller than the characteristic scale, over which $V_{\eff}$ is changing,  the effective potential can be approximated as a linear function of $(r-R_\res)$ around the resonance.
In this case the Landau-Zener formula appears
\begin{equation}
  P_{\rm na}= \exp\left[-\frac{\pi }{2 \gamma}\right].
\end{equation}
For small vacuum mixing angles one has
$\theta_{\rm in} \approx \frac{\pi}{2}$ and $\theta_{\rm out}\approx \theta_0\ll 1$, Eq.~\eqref{presfull} can be rewritten as
\begin{equation}
  P_{x\to s}= 1-\exp\left[-\frac{\pi }{2 \gamma}\right].
  \label{pres}
\end{equation}
Figure~\ref{EnergyLevels} illustrates the above considerations. Energy levels
$E_a(r)$ and $E_b(r)$ do not cross. The value $E_a-E_b$ reaches its minimum as
$r \to R_{\res}$. In the case of fully adiabatic propagation (\textit{i.e.}\ change of the radius)
one remains on the same energy level $E_a(r)$ or $E_b(r)$.  As a result, a
state $\ket{\nu_x}$ that is \textit{mostly $\ket{\nu_a}$} deep inside the star
would remain \emph{mostly $\ket{\nu_a}$} everywhere and would exit the star as
mostly sterile state $\ket{\nu_s}$.  
The probability of such a process for
$\theta_0 \ll 1$ is given by $P_{x\to s}^{\rm adiab} \sim \cos^2\theta_0\to 1$ --
the result familiar from the MSW effect in the Sun. 
This can be seen from Eq.~\eqref{pres} when the parameter of non-adiabaticity $\gamma \to 0$.

The non-diagonal elements in the Hamiltonian make propagation non-adiabatic.
Therefore, although levels do not cross, when they are approaching close to
each other, a transition between them can occur.  As a result, the probability
for an active neutrino to pass a resonance region without conversion remains
finite. One can consider the limit $\gamma \gg 1$, where the probability behaves as $P_{x\to s} \approx \frac{\pi}{2\gamma} = \frac{\pi^2}{2} \frac{R_\width}{L_\osc} \ll 1$.

\subsection{Mixing with the electron flavor}
\label{sec:mixing-with-nu_e}

The described mechanism can of course be used for $\nu_e - \nu_s$ mixing as
considered in a number of papers
\cite{Kainulainen:1990bn,Shi:1993ee,Nunokawa:1997ct,Tamborra:2011is}.  The
effective potential for $\nu_e/\bar\nu_e$ is, however, different
from~\eqref{Veff}:
\begin{equation}
  V_{\eff}^{\nu_e,\bar\nu_e}(r) = \mp\frac{G_F}{\sqrt{2}}N_{\rm b}\Bigl(-2Y_{e}+Y_n - 4Y_{\nu_e} - 2Y_{\nu_\tau} - 2Y_{\nu_\mu}\Bigr).
  \label{Veff2}
\end{equation}
(the upper sign is for $\nu_e$, the lower -- for $\bar\nu_e$).
Using the relations~\eqref{eq:beta-eq} one can see that the effective potential~\eqref{Veff2} changes its sign as one moves away from the core.
As a result, the production is possible for both electron neutrinos and antineutrinos.
While the resonant conversion for $\bar\nu_e$ proceeds similarly to $\bar\nu_x$, for electron neutrinos the resonance condition is satisfied at two different radii.
So $\nu_s$ converted at an inner radius can be re-converted to active neutrinos at an outer radius, reducing the effectiveness of the production (see e.g.~\cite[Fig.~3]{Nunokawa:1997ct}).
The kinetic equation~\eqref{kinetic_res} does not take this into account.
Another important effect is that the value of $Y_{\nu_e}$ is tightly connected with the electron-positron asymmetry $Y_e$ via beta-equilibrium condition.
Therefore, efficient resonant conversion may shift beta-equilibrium and in this was significantly affect the nucleosynthesis in supernovae (see \cite{Bliss:2018nhk}).
A proper self-consistent treatment of these processes are beyond the scope of this paper, therefore, we limit ourselves only to the mixing with $\mu$ and $\tau$ flavors.

\section{Back-reaction of sterile neutrinos}
\label{sec:Back-reaction}

\subsection{Evolution of $x$-flavor population}
\label{sec:evolution-x-flavor}

The active-sterile conversion depletes the number of anti-neutrinos of given energy at a given radius (the two are related via Eq.~\eqref{ResonanceEnergy_res}).
Therefore, the conversion could have led to the deviation of the $\bar\nu_x$ distribution function from its initial equilibrium form.
However, other processes such as nucleon-neutrino scatterings
\begin{equation}
  \bar\nu_x + N \to \bar\nu_x + N
  \label{eq:N-nu-scat}
\end{equation}
lead to the change of the shape of the anti-neutrino distribution function without changing the total number of anti-neutrinos at the radius $r$.
The nucleon-nucleon bremsstrahlung production of neutrino pairs,
\begin{equation}
  N+N \to N + N + \bar \nu_x + \nu_x,
  \label{eq:bremsstrahlung}
\end{equation}
partially re-populates the number of $\bar\nu_x$ (without changing the total lepton number).
The process~\eqref{eq:bremsstrahlung} is stopped by the \emph{neutrino} Pauli blocking.
The reaction rates of the processes~\eqref{eq:N-nu-scat}--\eqref{eq:bremsstrahlung} are faster than sterile neutrino conversion rate~\cite{Thompson:2000gv}.
Therefore we can always describe the population of \nux\ by the equilibrium distribution function,
\begin{equation}
  \label{eq:FD}
  \bar f_{x}(E,r,t) = \frac{1}{(2\pi)^3}\frac{1}{\exp\left[\frac{E+\mu_x(r,t)}{T(r)}\right]+1}
\end{equation}
(with $\mu_x \to -\mu_x$ for neutrino distribution function).
\emph{The evolution of the neutrino population is fully  encoded into the evolution of the chemical potential $\mu_x$.}\footnote{Recall that we only analyse the duration of time $\tpb \sim 1$~sec and therefore neglect temporal change of the temperature profile.}

The evolution of the chemical potential affects the effective potential $V_{\eff}$ and, therefore, the resonance energy \eqref{ResonanceEnergy_res} via the change of the lepton number $Y_x$.
It can be seen from~\eqref{ResonanceEnergy_res} that with the growth of $Y_{x}$ the resonance energy increases so that the number density of active anti-neutrinos with energy $E \ge E_{\res}$ diminishes and as a result the production stops.\footnote{Sufficiently large asymmetry could cause the effective potential to change the sign and therefore cause conversion $\nu_x \to \nu_s$.
  Such a process would result in washing out of the asymmetry.
  We will see below that this does not happen for the realistic values of the parameters.
}

The non-zero chemical potential $\mu_x \sim T$  means that neutrino average energy increases.
For the muon flavor large values of $\mu_x(r,t)$, increase the number of neutrinos that can participate in the production of muons in reactions, like
\begin{eqnarray}
  \nu_{\mu} + n \to p + \mu^-
  \label{muonprod1}                                                \\
  \nu_{\mu} + e^- \to  \mu^- + \nu_e                               \\
  \nu_{\mu} + \bar\nu_e \to \mu^- + e^+
  \label{muonprod3}
\end{eqnarray}
leading to the non-negligible population of $\mu^-$.
Similar reactions are possible for anti-neutrinos and anti-muons, but the number density of $\bar\nu_{\mu}$ is extremely small in this regime, leading to negligible production of $\mu^+$.
So the muon lepton asymmetry will be stored not only in neutrinos but in
muons as well.
The population of $\tau^\pm$ leptons remains negligible because of their large mass.

\subsection{Diffusion}
\label{app:diffusion}

\begin{figure}[!h]
  \includegraphics[width=0.45\textwidth]{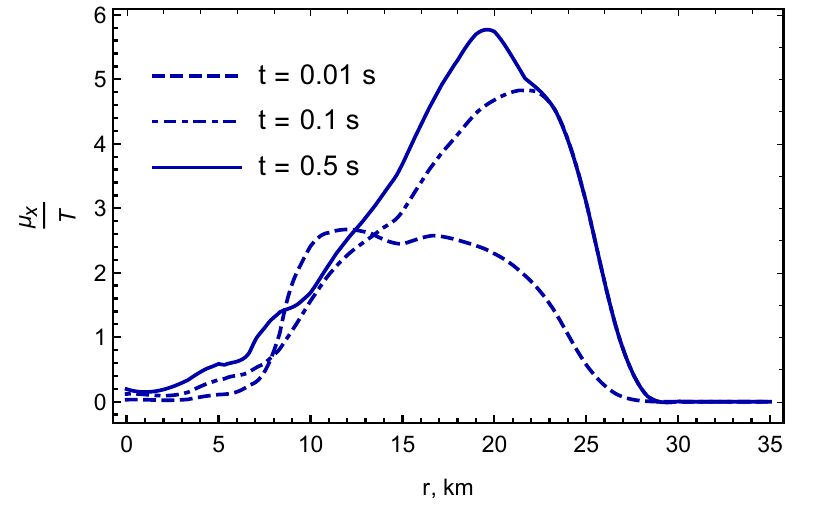} \hfill
  \includegraphics[width=0.45\textwidth]{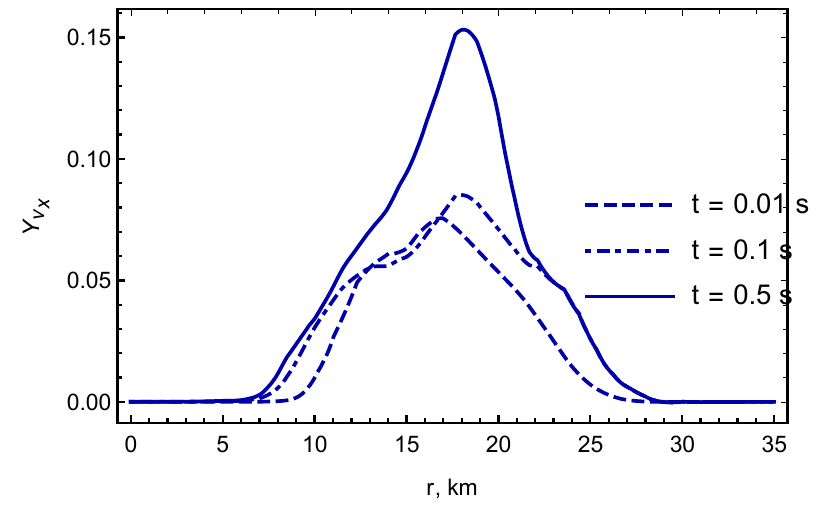}
  \caption{The same as for Fig.~ \ref{fig:x-asymmetry_profile}, but without diffusion.
    In this case, asymmetry increases at every point independently.
    In the absence of diffusion the maximum asymmetry is sufficiently larger, but  $V_{\eff}$ still does not change its sign. The peak position is changing slightly due to the change of resonance condition eq.~\eqref{ResonanceEnergy_res} with the build-up of the asymmetry $Y$ as well as with the change of parameters of SN.
  }
  \label{fig:x-asymmetry_profileNoDiff}
\end{figure}
The inhomogeneous chemical potential $\mu_x(r,t)$ triggers the lepton number diffusion processes. \emph{Neutrinos} (whose number exceeds greatly that of anti-neutrinos) diffuse away and the reactions like~\eqref{eq:bremsstrahlung} then replenish population of anti-neutrinos

A typical time scale for the diffusion over the distance $R$ is $t_{\rm Diff}=\frac{R^2}{\lambda_{\mfp}}$, where $\lambda_{\mfp}$ is the mean free path of (anti)neutrinos of $x$-flavor. 
The neutrino's mean free path depends on the neutrino energy and matter density.
A straightforward computation of neutrino scattering in a medium of non-relativistic nucleons gives $\lambda_{\mfp} \sim \frac{\pi}{G_F^2 N_{\rm b} E^2}$.\footnote{Recall that we are interested only in the diffusion of $\mu$ or $\tau$ flavors and therefore only neutral current processes contribute to the scattering of both neutrinos and anti-neutrinos.}
\textcolor{black}{
Typical values of neutrino energies in supernovae is $E \sim \mathcal{O}(100)$~MeV and densities can reach $N_{\rm b} \sim \unit[2\times 10^{38}]{cm^{-3}} $ so diffusion time can be as low as $\mathcal{O}(\unit[10^{-2}]{sec})$ -- much below the period of time over which we analyse the sterile neutrino production.
Therefore diffusion cannot be neglected. Strictly speaking, the diffusion approximation is not valid for $R\gtrsim R_{\nu\sph}$ where the density drops below $\rho <  \unit[10^{11}]{g/cm^{3}} $ and  neutrinos start to free stream. 
The region of the neutrinosphere thus serves as a ``sink'' of lepton asymmetry.
We can, however, ignore this correction thanks to the following consideration: \textit{(i)} neutrinos are actively converted at $R_\res \ll R_{\nu\sph}$. 
 \textit{(ii)} The value of the lepton asymmetry at $R \gg R_\res$ does not influence directly this conversion rate because the diffusion rate at these ``low'' densities becomes very high ($t_{\rm Diff} \ll 1$~sec). 
 As a result lepton asymmetry is washed out faster than it is produced.
 So it cannot accumulate and affect the value of the asymmetry in the inner region. 
 To check these arguments we artificially increased the diffusion coefficient at $R \sim R_{\nu\sph}$ to effectively mimic free-streaming of neutrinos. 
 The resulting asymmetry evolution appeared to be absolutely \textit{identical} to the original scenario at a given accuracy. 
 Therefore, no additional treatment for the lepton number inside the neutrinosphere is needed.}

To describe the evolution of the lepton asymmetry we use ~\eqref{asymmetryevolution}  with the diffusion coefficient $D(r,E)$ given by the relaxation time-approximation:
\begin{equation}
  D(r,E)=\frac{\lambda_{\mfp}(r,E)}{3} = \frac{\pi}{3G_F^2 N_{\rm b}(r) E^2}
  \label{eq:3}
\end{equation}
(Appendix ~\ref{asymmetryappendix}).

The collisional production of sterile neutrinos can also affect the evolution of the chemical potential. 
Indeed, let $\Gamma_{\nu_x\to \nu_s}^{\coll}$ be the rate of collisional production of sterile neutrinos $\nu_x \to \nu_s$, while $\Gamma_{\nux\to \nu_s}^{\coll}$ be a similar rate for anti-neutrino production
(of course, $\nu_x$ and $\nux$ produce sterile states of opposite helicity). 
Naively, one could argue that as there are more $\nu_x$ than $\nux$ in the resonance region, the collisions will predominantly convert  $\nu_x \to \nu_s$, thus decreasing the asymmetry. 
This is, however, not the case as the collision rates are not the same, $\Gamma_{\nu_x\to \nu_s}^{\coll}\ll \Gamma_{\nux\to \nu_s}^{\coll}$ in the resonance region, see \textit{e.g.}~\cite{Abazajian:2001nj} where the resonance enhancement/suppression of the collisional production rate is discussed. 
Indeed, the collision rates are proportional to $\sin^2\bigl(2\theta\bigr)$.
In the resonance region, angle for anti-neutrinos is $\theta_\res^\nux \sim \mathcal{O}(1)$, while for neutrinos $\theta^{\nu_x}_\res \simeq \frac 12 \theta_0$, as one can see by replacing $V_\eff \to -V_\eff$ in Eq.~\eqref{eq:mixing_angle} and making use of the condition~\eqref{rescondition}.
As a result
\begin{equation}
    \label{eq:coll_rates}
    \Gamma_{\nu_x\to \nu_s}^{\coll} \sim \theta_0^2 \Gamma_{\nux\to \nu_s}^{\coll}
\end{equation}
With chemical potential reaching $\mu_x/T \sim 3$ (see Fig.~\ref{fig:x-asymmetry_profile}) $n_{\nux} \sim 10^{-2} n_{\nu_x}$ and therefore we conclude that collisions do not contribute significantly to the wash out of lepton asymmetry for mixing angles that we are considering.

\subsection{Effects on the electron flavor population}
\label{sec:effects-nu_e}

As mentioned before, in the case of the mixing with $\nu_{\mu}$, the development of the chemical potential of the muon lepton number would lead to the asymmetry in charged muons.
In its turn, this will affect electrons and electron neutrinos via the charge neutrality condition $ Y_p = Y_e + Y_{\mu}$, changing the asymmetry of electrons.
Connection between charged leptons and correspondent neutrinos is expressed with beta-equilibrium relations:
\begin{equation}
  \mu_{\mu}-\mu_{\nu_\mu} = \mu_n-\mu_p = \hat{\mu}
\end{equation}
\begin{equation}
  \mu_{e}-\mu_{\nu_e} = \mu_n-\mu_p = \hat{\mu}
\end{equation}
Charge neutrality and beta-equilibrium allow us to connect all these parameters of the supernovae medium.
As a consequence of the increase of the muon neutrino chemical potential, we
will have decreased values of electron density and density of $\nu_e$.
This effect, however, affects the overall results only marginally, as even with back-reaction, muon neutrinos chemical potential (see Fig.~\ref{fig:x-asymmetry_profile})

\section{Lepton asymmetry evolution}
\label{asymmetryappendix}
We start from radial diffusion equation for distribution function with a source
\begin{equation}
  \frac{\partial f_x(r,E,t)}{\partial t} =\frac{1}{r^2}\frac{\partial}{\partial r}\left(r^2D(r,E)\frac{\partial f_x(r,E,t)}{\partial r}\right) + I_x(r,E,t)
    \label{diffrdial}
  \end{equation}
  where $f_x$ - distribution function of $\nu_x$ ($\bar\nu_x$), $D(E,r)$ -- diffusion coefficient, $I_x(r,E,t)$ -- source.
By taking Eq.~\eqref{diffrdial} for neutrinos and anti-neutrinos, integrating their difference over momentum, and dividing by $N_{\rm b}$ we find:
\begin{equation}
  \frac{\partial Y_x(r,t)}{\partial t} =  \frac{1}{N_{\rm b}(r)}\frac{1}{r^2}\int\frac{\partial}{\partial r}\left(r^2 D(r,E)\frac{\partial}{\partial r}( f_{x}(E,r,t)-\bar f_{x}(E,r,t))\right)d^3p + S_x(r,t)
  \label{asymmetryevolution}
\end{equation}
here $S_x(r,t)$ is the integrated source of asymmetry
\begin{equation}
  S_{x}(r,t) = \frac{\pi}{N_{\rm b}(r)} E_{\res}^2(r,t) \bar{f^{out}_x}(E_{\res}(r),r,t)P_{x\to s}(E_{\res}(r),r,t) \frac{dE_{\rm res}}{dr}(r,t) 
\end{equation}
Combining these results together, we arrive to the final equation describing the evolution of lepton number, Eq.~\eqref{diffsimplified_res}.

\section{Quantifying the uncertainties}
\label{app:uncertainties}

The production of sterile neutrinos is most sensitive to the maximum temperature in the SN  as it defines the population of the highest-energy active neutrinos, that will be available for conversion. 
In order to quantify the uncertainties due to variation of different parameters, we adopt the toy model which has no temporal evolution. In this way, we can estimate the sensitivity of our results on the models not measured directly.

The baryon density is approximated as a constant inside the supernova core ($r < R_{\rm core}$) and decays exponentially at larger radii,
\begin{equation}
\label{eq:profile}
    \rho_B=\rho_0\exp\left[-\frac{r-R_{\rm core}}{R_{\rm core}}\right],\quad r >
R_{\rm core}
\end{equation}
Temperature is chosen to decrease linearly from $T_{\rm max}$ at $r = 0$ to $T_{\rm min}$ at $r = 50 \km$ and is also constant during the first second.
Proton number fraction remains constant and it is
just a simplification for our model (note that does not necessarily mean that we define the number of electrons as there may be a change of population of other charge massive leptons). 
Numerical values of the relevant parameters are specified in Table~\ref{tab:SNparam}.

Comparison with the simulation snapshots (Fig.~\ref{fig:SN_model}) shows, that the values of asymmetries are on the same order of value, while density decreases slower and temperature can be both higher, and lower, than in the fiducial model but is, again of the same order. So, our toy model serves as a fair representation of the realistic profile.
\begin{table}[!t]
  \begin{center}
    \begin{tabular}{ | l | l |}
      \hline  
      Core radius                                                          & $R_{\rm core} = 10 \km$                              \\ \hline
      Max. Temperature                                                     & $T_{\rm max} = 30 \MeV$                              \\ \hline
      Min. Temperature                                                     & $T_{\rm min} = 3 \MeV$                               \\ \hline
      Baryon core density                                                  & $\rho_{0} = 3 \times 10^{14} \frac{\rm g}{\rm cm^3}$ \\
      \hline
      Baryon core number density                                           & $N_{0} =\unit[ 10^{38}]{cm^{-3}} $                   \\
      \hline
      Proton fraction                                                     & $Y_{p} = 0.3$                                        \\ 
      \hline
    \end{tabular}
  \end{center}
  \caption{Parameters of the toy model of the supernova adopted in this section.
    Temperature is chosen to decrease linearly from $T_{\rm max}$ at $r = 0$ to $T_{\rm min}$ at $r = 50 \km$ and is also constant during the first second.
    See Appendix~\protect\ref{app:SNmodel} for other details.}
  \label{tab:SNparam}
\end{table}
\begin{figure}[!t]
  \centering
  \resizebox{\textwidth}{!}{\includegraphics[height=3cm]{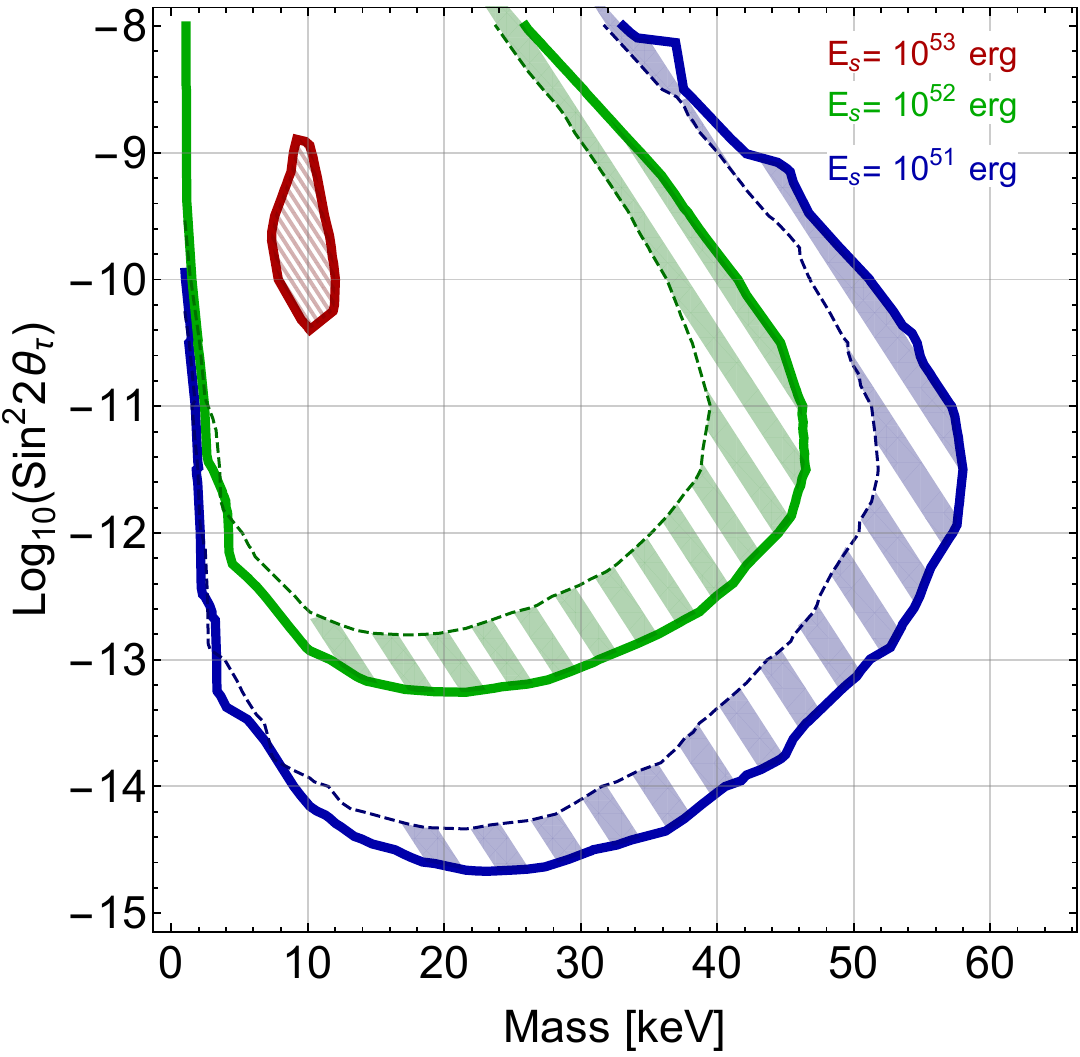}~~
    \includegraphics[height=3cm]{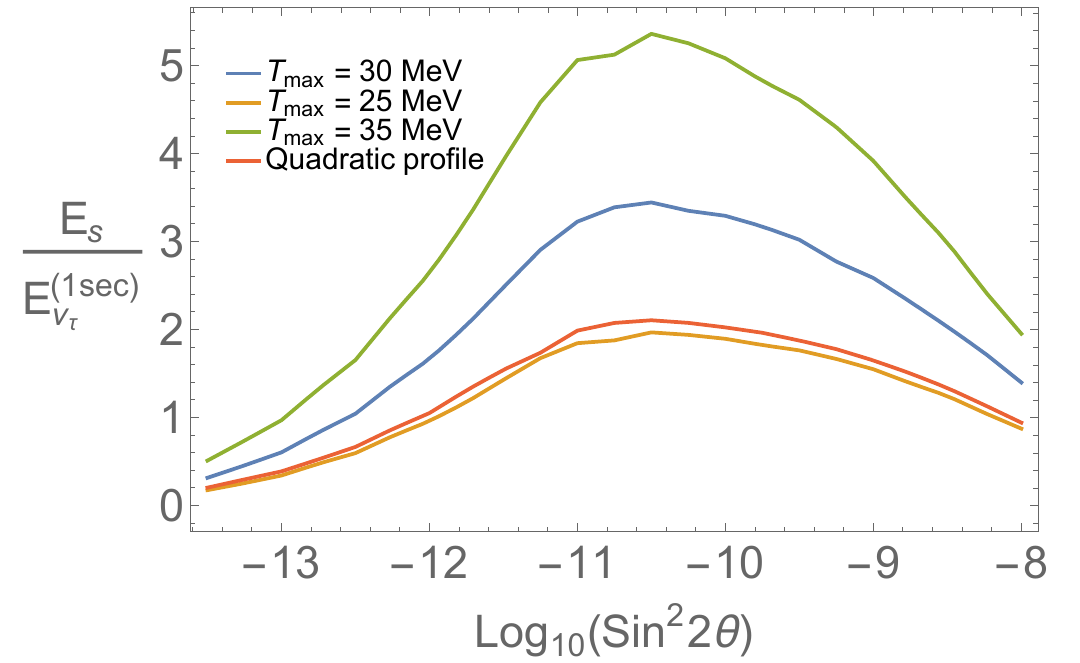}}
  \caption[Energy emitted in the toy model and its uncertainty]{
  \textit{Left panel:} Energy emitted by sterile neutrinos in toy model (thick lines) and modified toy model of  Section~\ref{app:uncertainties} (thin dashed lines) - when central temperature in the SN is decreased to 25 MeV, which correspond to scaling of temperature for $\approx$ 15\%. Shaded region shows the corresponding "uncertainty" of production. Sterile neutrino are considered mixed solely with $\nu_\tau$.
  \textit{Right panel:} Uncertainties related to the SN temperature models. Energy, emitted in the form of sterile neutrinos as a function of the mixing angle for the mass $m_s = 20$~keV.
The curves show the effects of changing the maximal temperature $T_{\rm max}$ by $\pm 5$~MeV as well as and different scaling of the temperature profile between $T_{\rm max}$ and $T_{\rm min}$ (quadratic rather than linear).
  }
  \label{fig:energy_emitted}
\end{figure}

We see, that although the parameters of the SN in specific regions differ significantly (at the outer radii $r \gtrsim 20-30$ km for density and inner radii $r \simeq 10$ km for temperature), the maximum total energy outcome has not changed significantly. This happens due to a very strong back-reaction that localizes the production to a compact spatial region in the interior of the neutron star.

In order to quantify the uncertainties, 
The change of energy outcome with temperature decrease is also of the same order (while the temperature modification has also a similar factor of 15-20 \%). It shows, that uncertainty of the result due to exact temperature inside supernovae was not a feature of the specific model we used. 

\clearpage

\bibliography{Biblio/preamble,Biblio/SNbib}
\end{document}